\documentclass[a4paper,fleqn,usenatbib]{mnras}

\usepackage[T1]{fontenc}
\usepackage{ae,aecompl}
\usepackage{multicol}

\usepackage{graphicx}	% Including figure files
\usepackage{amsmath}	% Advanced maths commands
\usepackage{amssymb}	% Extra maths symbols
\usepackage{float}
\usepackage{array}
\usepackage{placeins}
\usepackage[table]{xcolor}
\usepackage{multirow}
\usepackage{bigstrut}
\usepackage{siunitx}
\usepackage{soul}
\graphicspath{{./}}

\def\xmm{\textit{XMM-Newton} }
\def\xmmnogap{\textit{XMM-Newton}}
 
\def\swift{\textit{Swift} }
\def\msun{$M_{\odot}$}
\makeatletter
    \def\fps@figure{h!tbp}
    \def\fps@table{h!tbp}
\makeatother
\DeclareSIUnit\angstrom{\text {Å}}

\title[Testing disc reprocessing with optical PSD of NGC 4395]{Testing disc reprocessing models for AGN optical variability by comparison of X-ray and optical power spectra of NGC~4395}

\author[Beard, M. W. J. et al.]{
M. W. J. Beard,$^{1}$\thanks{E-mail: mb15g14@soton.ac.uk (MWJB)}
I. M. M$\rm^{c}$Hardy,$^{1}$,
K. Horne$^{2}$,
E. M. Cackett$^{3}$,
F. Vincentelli$^{1,4,5}$, \and
J. V. Hern\'andez Santisteban$^{2}$,
J. Miller$^{3}$,
V. S. Dhillon$^{6,5}$
J. H. Knapen$^{5,4}$
S. P. Littlefair$^{6}$ \and
D. Kynoch$^{1}$, 
E. Breedt$^{7}$,
Y. Shen$^{8}$
J. Gelbord$^{9}$
\\
\\
$^{1}$ University of Southampton, University Road, Southampton SO17 1BJ, UK\\
$^{2}$ School of Physics and Astronomy, University of St Andrews, North Haugh, St Andrews KY16 9SS, UK\\ 
$^{3}$ Department of Physics and Astronomy, Wayne State University, 666 W.
Hancock Street, Detroit, MI 48201, USA\\
$^{4}$ Departamento de Astrof\'{i}sica, Universidad de La Laguna, E-38206 La Laguna, Spain\\
$^{5}$ Instituto de Astrof\'{i}sica de Canarias, V\'{i}a L\'{a}ctea S/N, E-38205 La Laguna, Spain \\
$^{6}$ Dept of Physics and Astronomy, University of Sheffield, Sheffield S3 7RH\\
$^{7}$ Institute of Astronomy, Madingley Road, Cambridge, CB3 0HE\\
$^{8}$ Department of Astronomy, University of Illinois at Urbana-Champaign, 1002 W. Green Street, Urbana, IL 61801, USA\\
$^{9}$ Spectral Sciences Inc, 4 Fourth Avenue, Burlington, MA 01803, USA
}
\date{Accepted XXX. Received YYY; in original form ZZZ}

\pubyear{2022}

\begin{document}
\label{firstpage}
\pagerange{\pageref{firstpage}--\pageref{lastpage}}
\maketitle

\begin{abstract}
It is generally thought that AGN optical variability is produced, at least in part, by reprocessing of central X-rays by a surrounding accretion disc, resulting in wavelength-dependent lags between bands. Any good model of AGN optical variability should explain not only these lags, but also the overall pattern of variability as quantified by the power spectral density (PSD). Here we present $\sim$daily g'-band monitoring of the low-mass AGN NGC\,4395 over 3 years. Together with previous TESS and GTC/HiPERCAM observations we produce an optical PSD covering an unprecedented frequency range of $\sim7$ decades allowing excellent determination of PSD parameters. The PSD is well fitted by a bending power law with low-frequency slope $\alpha_{L} = 1.0 \pm 0.2$, high-frequency slope $2.1^{+0.2}_{-0.4}$ and bend timescale $3.0^{+6.6}_{-1.7}\,$\,d.
This timescale is close to that derived previously from 
a damped random walk (DRW) model fitted  to just the TESS observations, although $\alpha_{L}$ is too steep to be consistent with a DRW. We compare the observed PSD with one made from light curves synthesized assuming reprocessing of X-rays, as observed by \xmm and Swift, in a disc defined by the observed lags.
The simulated PSD is also well described by a bending power law but with a bend two decades higher in frequency. We conclude that the large-amplitude optical variations seen on long-timescales are not due to disc reprocessing but require a second source of variability whose origin is unknown but could be propagating disc accretion rate variations.

\end{abstract}


%%%%%%%%%%%%%%%%%%%%%%%%%%%%%%%%%%%%%%%%%%%%%%%%%%

%%%%%%%%%%%%%%%%% BODY OF PAPER %%%%%%%%%%%%%%%%%%

\section{Introduction} \label{intro}

Active Galactic Nuclei (AGN) have long been observed to emit X-ray continua that vary rapidly and non-periodically \citep[e.g.][]{mchardy_1987,lawrence_1987}, consistent with the idea that the X-ray emission originates from a compact region within a few Schwarzschild radii of the central black hole. X-ray variability is often characterised by the power spectral density (PSD) of the lightcurve. From early EXOSAT observations it was shown that AGN X-ray PSDs are well described by bending power laws, $P({\nu}) \propto \nu^{-\alpha}$, with $\alpha \sim 2$ at high frequencies ($\geq 10^{-5}$\,Hz), flattening to $\alpha \sim 1$ below a bend frequency $\nu_{B}$ \citep{mchardy_1988}. By comparison with the PSDs of Galactic X-ray binary systems  \citep[e.g.][]{nolan_1981} it was proposed that the bend timescale was proportional to the black hole mass \citep{mchardy_1988}, a proposition which was supported by many subsequent observations \citep[e.g.][]{nandra_2001, uttley_2002, mchardy_2004, mchardy_2006, ponti12, tortosa23, serafinelli24}. 
 
The origin of short-timescale (hours / days) optical variability in AGN was a mystery for many years. However, many observations over the last 2 decades have shown a strong correlation between X-ray and optical variations, with the optical lagging behind the X-rays. This correlation strongly suggests that at least part of the optical variability is driven by reprocessing of X-rays. There is good evidence that the accretion disc, the broad line region (BLR) gas and quite possibly also a disc wind act as reprocessors although there is still considerable discussion as to their relative importance, e.g. \cite{korista_goad_2001,korista_goad_2019,uttley_2003,cackett_2007_testing, arevalo_2008_correlated,breedt_2009,breedt_2010,shappee_2014,mchardy_2014_swift,mchardy_2016_the,mchardy_2018,edelson_2015_space,edelson_2019, cackett_2018,netzer_2019, cackett_2020, vincentelli_2021,vincentelli_2022,kara_2021,hagen24} and many others. 

If the optical variability of AGN is mostly due to reprocessing of X-rays, then the characteristic signature of the X-ray variations, i.e. the shape of the X-ray PSD, should be present in the PSD of the optical variations. In particular, the optical PSD should be the product of the PSD of the X-ray variations with the square of the modulus of the response function of the reprocessing material \citep[e.g. see][]{arevalo_2009}.
Alternatively, the optical PSD can be produced by convolving the X-ray lightcurves with the response function and then calculating the PSD of the resulting reprocessed optical lightcurves.
Comparison of an observed optical PSD with a model PSD based on reprocessing is therefore a powerful tool for testing whether the reprocessing model is correct and whether it explains all optical variability. This comparison can be carried out independently of measurement of an interband lag. However, if combined with measurement of a lag, or multiple interband lags, which constrain the response function, the test is stronger \citep{arevalo_2009,panagiotou_2022,hagen23}. Comparison of the optical and X-ray PSDs of one particularly well-studied AGN is the aim of the present paper.

AGN optical PSDs have been studied previously by a number of previous authors. We discuss some of the main observations here.
\cite{czerny_1999} modelled the optical PSD of NGC~5548 with a broken power law, claiming a break at $\sim100$\,d. However they did not take into account PSD distortions from red noise leak and irregular sampling so the timescale uncertainty is unclear. 
\cite{arevalo_2009} measured the X-ray and optical power spectra of NGC~3783. They found that the X-ray PSD was described by a bending power law with a bend at $\sim3$\,d. However the optical PSD continued rising as an unbending power law towards the longest timescales covered ($\sim300$\,d). Thus the X-ray power exceeded the optical power at high frequencies but vice-versa at low frequencies. \cite{arevalo_2009} also measured the lag of the X-rays by the optical B and V bands ($\sim$6d) and constructed response functions for a variety of reprocessing geometries to determine which one best reproduced the observed ratio of the optical to X-ray PSDs. Given the 6d lag, no geometry could smooth the X-ray variations sufficiently to produce the observed optical variability, particularly at low frequencies.

\cite{breedt_2010} carried out a detailed X-ray and optical variability study of NGC~4051 using lightcurves of over 12 years.
They found that, while the X-ray PSD was well described by a bending power law \citep{mchardy_2004}, the optical PSD was adequately described by an unbending power law of slope $\sim1.4$, intermediate between the high-frequency ($\sim$2) and low-frequency ($\sim1.1$) X-ray PSD slopes. This intermediate slope may have resulted from the optical data not being good enough to properly define a PSD bend.

\cite{breedt_2010_thesis} calculated the optical PSDs of 6 more bright Seyfert 1 Galaxies (NGC~3227, NGC~4593, NGC~5548, NGC~7469, Mrk79, Mrk110) and compared the optical PSDs with the X-ray PSDs derived by \cite{summons_2007}. 
The results were broadly similar to those of \cite{arevalo_2009} finding, in three cases (NGC 5548, NGC 4593 and Mrk 79), that the optical power exceeded the X-rays at low frequencies. Although bending powerlaws could be fitted to the optical PSDs, in no case were they a better fit than an unbending power law.

The optical data used by the above authors, although covering periods of more than a decade, were generally limited to ground-based observations once per night. Considerably improved sampling was provided by the Kepler \citep{borucki_2010} and TESS \citep{ricker_2014} space observatories. Kepler observed selected areas continuously in white light with 30min sampling for one or more periods of 3-months (quarters). TESS also has 30min sampling with continuous observations of $\sim1 $month in a broad red (600-1000nm) observing band.

From Kepler data \cite{mushotzky_2011} measured the PSDs of 4 type 1 AGN, finding unbending power law slopes between 2.6 and 3.3. \cite{smith_2018} found a similar single slope range for a sample of 21 type 1 AGN although, for the 6 PSDs with the steepest high-frequency slopes ($\alpha_{H}$), bending power laws were a better fit. The bend timescales (9 to 53d) roughly scaled with mass.
\cite{smith_2018} do not discuss the slopes below the bend ($\alpha_{L}$) but visual inspection of their Fig.10 indicates $\alpha_{L} > 0$ in four AGN and $\alpha_{L} \leq 0$ in two.
Probably the highest S/N Kepler AGN PSD is that of Zw 229-15 \citep{edelson_2014}. They find a bending PSD with a very steep high-frequency slope ($\sim3.6$) flattening, on timescales $>5$\,d, to $\alpha_{L} \sim2$.  Unfortunately, there is no X-ray PSD available for comparison. \citep{edelson_2014} also found no evidence for rms/flux relationships or lognormal flux distributions, which are common in X-ray AGN lightcurves, and so they concluded that X-ray reprocessing did not drive the optical variability.

The PSD slopes from \cite{mushotzky_2011} are interesting because they are steeper than the value of 2 expected for the high-frequency PSD slope in the Damped Random Walk (DRW) model which is commonly applied to AGN optical variations \citep[e.g.][]{kelly09}. The DRW PSD is also characterised by a low-frequency PSD slope of zero and a damping timescale, $\tau$, equivalent to the bend timescale of a bending power law PSD, which \cite{burke21} have shown scales with mass from white dwarfs to AGN. Using variants of the DRW model, \cite{zu13_drw}, \cite{stone22} and \cite{yu22_beyond_drw}, all suggest a steepening of the PSD slope at the highest frequencies, possibly indicating a second, high-frequency, bend timescale.

Measurement of low-frequency PSD slopes in bending or breaking power law models is rare. However, \cite{simm16} using continuous-time autoregressive moving average (CARMA) models \citep{kelly14}, found slopes consistent with 1 in the PSDs derived from Pan-STARRS1 lightcurves of AGN in the Cosmos-XMM field. Above the bend they found a range of slopes between 2 and 4. These low-frequency slopes again cast doubt on at least the simple version of DRW as the explanation of all AGN optical variability.

Of particular relevance to the present paper, we note the results of \cite{burke_2020} on NGC~4395. They model TESS data with a DRW, finding a damping timescale of ~2.3d. They also fit a broken power law to the PSD covering 2 decades in frequency from $\sim3 \times 10^{-7}$ to $\sim3 \times 10^{-5}$ Hz, finding $\alpha_{H} \sim2$ and $\alpha_{L} \sim0$. The damping timescale is towards the bottom end of the timescales covered by TESS and a major aim of the present paper is to refine these results with larger frequency coverage.

\cite{panagiotou_2022}, broadly following \cite{arevalo_2009} and \cite{breedt_2010}, modelled together the multiband PSDs and lags of NGC~5548 using data from Swift \citep{edelson_2015_space}. Covering a slightly lower frequency band than TESS, and with reduced frequency range ($\sim1.5$\,decades), they found power law PSD slopes $\sim 2$. 
Unlike \cite{arevalo_2009} they concluded that X-ray disc reprocessing could adequately describe both the lags and the PSDs. 

Although reprocessing of central X-rays provides a good explanation of short-term optical variability, there are longer-term (months/years) variations in the optical lightcurves which often have no parallel in the X-ray band \citep[e.g.][]{arevalo_2008_correlated, breedt_2010}. These variations may explain why, at the very lowest frequencies, the optical PSDs of AGN often have more power than the X-ray PSDs.
The origin of the long-timescale variations is unknown but they may represent intrinsic variability of the quiescent accretion disc, perhaps as a result of inwardly propagating accretion rate variations \citep{arevalo_2006_investigating}. Such variability would occur not on the fast light-travel timescales of reprocessing but on the much slower viscous timescale. Slow moving temperature fluctuations, travelling both inward and outward in the disc, have recently been found by \cite{neustadt22} and \cite{stone23} from analysis of Swift UV/optical and SDSS lightcurves respectively. The physical origin of these fluctuations is not yet clear but they may be related to the non-reverberation signals seen by, for example, \cite{arevalo_2008_correlated} and \cite{breedt_2010}. 

The situation, then, regarding how much of the optical variability of AGN can be attributed to reprocessing of X-rays and, if so, reprocessing from what, is not yet entirely clear. 
Whether optical AGN PSDs can be explained easily by the disc reprocessing scenario is even less clear. To provide a strong test of the reprocessing scenario for PSDs there are a number of requirements. Firstly we require very good X-ray and optical lightcurves so that we can produce high S/N PSDs in both bands. Secondly, we require frequency coverage over a large range, at least 4 decades in X-rays, so that we can properly define the PSD shapes and can measure accurately $\alpha_{H}$, $\alpha_{L}$ and $\nu_{B}$. In the optical band we require even more. We need to be able to measure those same three parameters over the frequency range where we expect reprocessed X-rays to be found. However we need to extend to even lower frequencies to test whether there are any other components in the optical variability, eg the intrinsic disc variations suggested by \cite{arevalo_2009} and \cite{breedt_2010}. 

The above requirements limit the number of AGN that can be observed in a human lifetime. In particular, we require a low-mass AGN where timescales are shorter and more accessible. The best candidate is NGC~4395.
Although its exact mass is still under debate, there is agreement that it is very low. From UV line reverberation mapping, generally considered to be one of the better methods for measuring AGN masses,
\cite{peterson_2005} derive a mass of $3.6 \times 10^{5}$ \msun and, from gas dynamical modelling, \cite{denbrock15} derive $4 \times 10^{5}$ \msun. However, from optical photometric reverberation, aimed mainly at $H_{\alpha}$, \cite{edri12_4395} derive $4 \times 10^{4}$ \msun, \cite{woo19_4395} derive $1 \times 10^{4}$ \msun and \cite{cho21_4395} derive $1.7 \times 10^{4}$ \msun. The exact value is not critical to the analysis presented here as the median point of the g'-band response function, which we use later to synthesise an optical light curve based on an X-ray light curve, is fixed by observations of the X-ray to g'-band lag \citep{mchardy_2023}. 
Thus, if we chose a low mass, we would simply increase the accretion ratio to produce a response function broadly similar to that of a higher mass and lower accretion ratio.

NGC~4395 is also well observed in both X-ray and optical/UV bands. 
The X-ray PSD is well defined from \xmm observations. From a single orbit ($\sim$120ks) observation \cite{vaughan_2005} obtained a good fit with a breaking power law with $\alpha_{H} = 1.92^{+0.28}_{-0.24}$, $\alpha_{L} =  1.2^{+0.76}_{-0.4}$ and $\nu_{B} = 1.9^{+0.8}_{-1.5} \times10^{-3}$ Hz. Thus, the break, or bend, timescale is a very accessible $\sim500-1000$s. 

We have subsequently observed NGC~4395 for a further 4 \xmm orbits, (Vincentelli et al and Beard et. al, in prep) which broadly confirm these PSD parameters. In this paper we use these new X-ray data, together with X-ray data covering longer timescales from Swift, to simulate optical lightcurves, assuming disc reprocessing, and hence to produce model optical PSDs. These model PSDs are then compared with the real, observed, optical PSD to determine whether disc reprocessing can account for all of the observed optical variability.

To produce the observed optical PSD the highest frequencies, $4 \times 10^{-4}$ to $\sim 10^{-1}$\,Hz, are provided by recent very high quality simultaneous $ugriz$ monitoring with 3s-sampling over 6hr using HiPERCAM on the 10.4m Gran Telescopio Canarias (GTC) \citep{mchardy_2023}. 
Intermediate frequencies (approx $5 \times 10^{-7}$ to $4 \times 10^{-4}$\,Hz) are provided by TESS \citep{burke_2020}.
To cover the lowest frequencies, $\sim2 \times 10^{-8}$ to $2 \times 10^{-6}$\,Hz,
we combine these data with nightly g'-band observations over a period of 3 years with a number of ground-based telescopes, including the Liverpool Robotic Telescope (LT), the Las Cumbres Observatory (LCOGT), the Zwicky Transient Factory (ZTF) and the Zowada Observatory. 
We also include V-band data with 2d sampling from two 4 month observation periods with Swift \citep{cameron_2012_correlated}.
Thus, overall, we are able to determine the optical PSD from $\sim 2 \times 10^{-8}$ to $\sim 10^{-1}$\,Hz.

The X-ray and optical observations are presented in Section~\ref{obs}. In Section~\ref{method} we derive the 
synthetic optical lightcurves expected from disc reprocessing of the new \xmm X-ray lightcurves. In Section~\ref{Analysis}
we derive both the observed and synthetic optical PSDs and in Section~\ref{conclusion} we compare these PSDs and 
discuss the implications of the comparison.

\section{Observations} \label{obs}
In this Section we present the observations that will be used later in the paper.

\subsection{\xmm Observations}

\begin{figure}
    \includegraphics[width=\columnwidth]{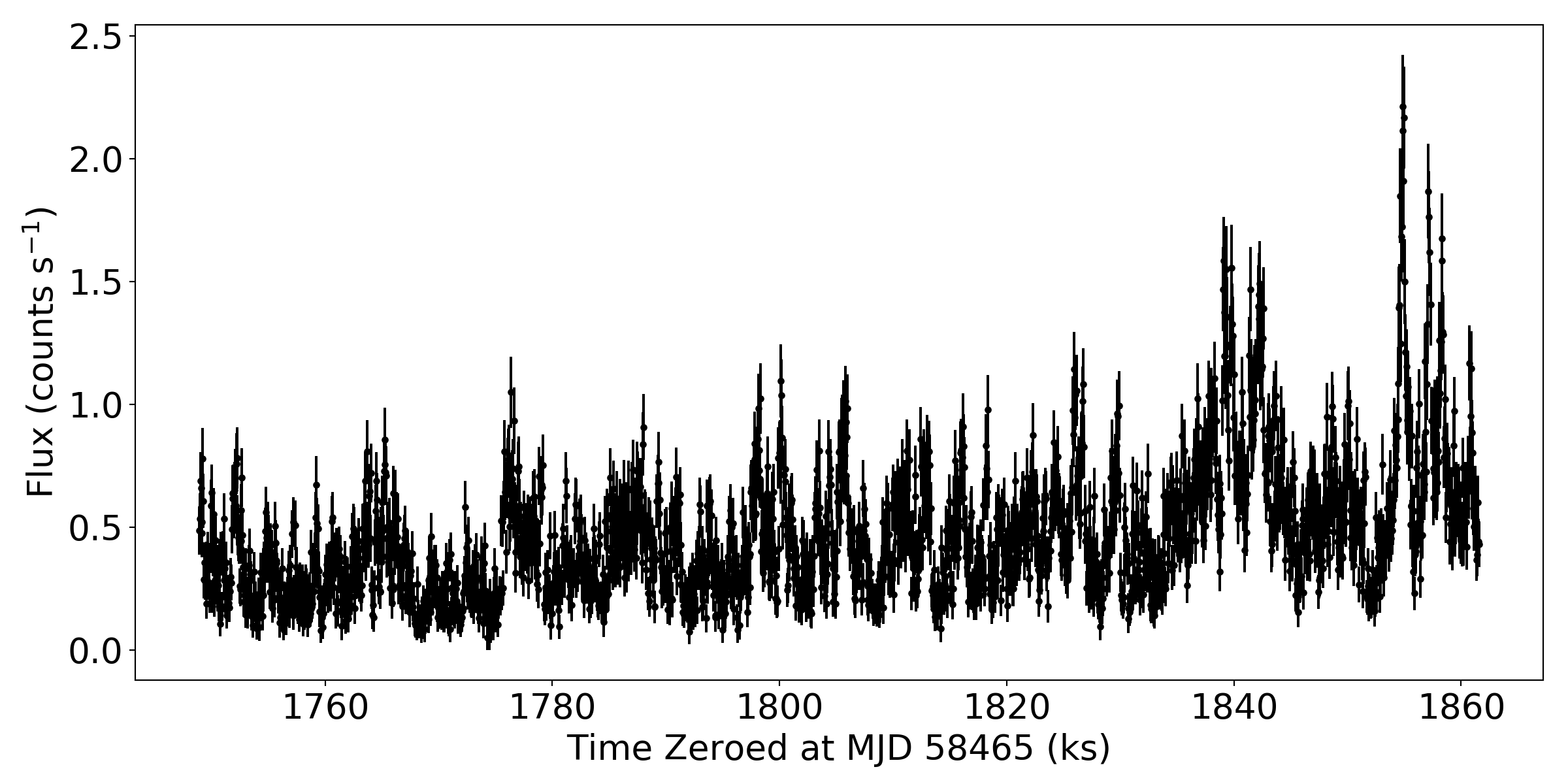}
    \caption{\xmm 0.5-10keV light curve for NGC 4395 using data from the night of January 2nd 2019, binned up to 50s.}
    \label{fig:XrayLC}
\end{figure}

Four separate $\sim$117 ks X-ray observations from the \xmm EPIC PN were taken, beginning on the 13th, 19th, and 31st of December 2018 and 2nd of January 2019. The data were analysed in the standard manner using the ESA XMM-SAS 19 software. We extracted source plus background events in 10s time bins in the energy band 0.5-10 keV from a circle of radius 30 arcsec around the AGN, filtering events with  \textsc{PATTERN<=4} and \textsc{flag==0}. The background was taken from a nearby blank area of the detector of the same size. As an example the light curve from the fourth observation, binned up to 50s, is plotted in Fig.~\ref{fig:XrayLC}.
\par

\subsection{Optical Observations}

For a valid comparison of the observed and synthetic optical PSDs, both PSDs must be in the same waveband. Here we choose the g'-band for a number of reasons. Almost all ground-based telescopes have g'-band filters and detectors are also quite sensitive in the g'-band, thus good quality g'-band lightcurves can be obtained reasonably easily. Although fractional variability is higher in the u-band, telescope sensitivity is considerably less and atmospheric absorption is more of a problem, so it is very hard to obtain long, high-quality, u-band observations. The g'-band is a good compromise. For some observatories, particularly TESS, only broad-band observations are available. In that case we adjust the variations in the TESS lightcurve to those expected in the g'-band. Here we present new g'-band observations from the LT, LCOGT, Zowada and ZTF and also show previously published optical lightcurves from Swift and TESS.

\subsection{Short Timescales: seconds - hours}
\label{short}

\begin{figure}
    \includegraphics[width=\columnwidth]{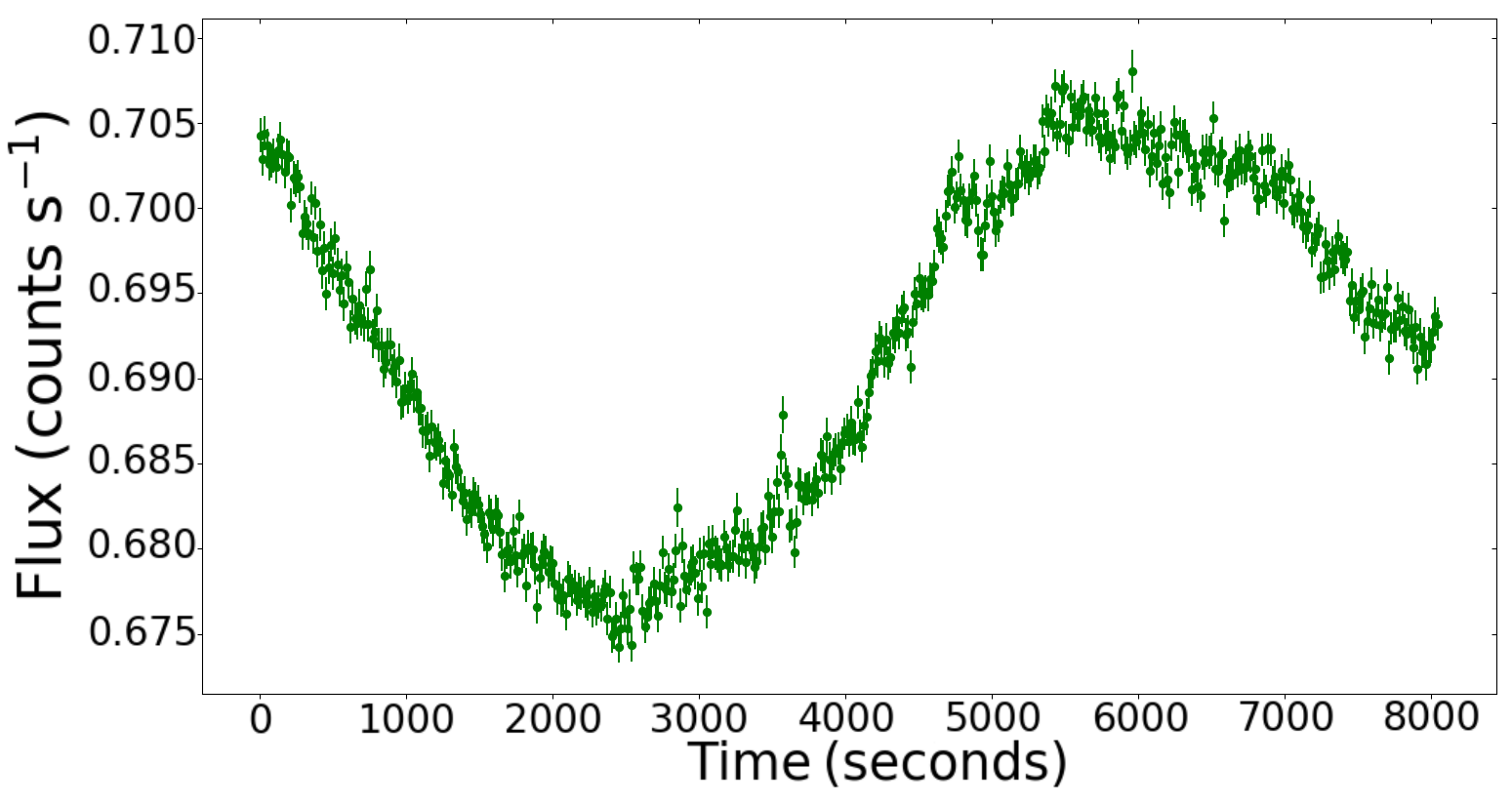}
    \caption{GTC HiPERCAM SDSS g' band light curve zeroed to MJD 58224.87378.
    }
    \label{fig:HCMLC}
\end{figure}

\underline{\bf HiPERCAM:~~}
\cite{mchardy_2023} carried out very high S/N simultaneous {\it ugriz} multiband monitoring of NGC~4395 with HiPERCAM \citep{dhillon21} on the 10.4m Gran Telescopio Canarias (GTC)\footnote{\url{http://www.gtc.iac.es/}} for 22ks continuously on the night of 16-17 April 2018.
In Fig.~\ref{fig:HCMLC} we show the first 8000s of the HiPERCAM $g'$-band light curve, with 15s sampling. Tracking problems slightly affected the later parts of the lightcurve so we do not use them here. As we are using these data to measure very high frequencies, slight loss of lower frequencies, which are covered by TESS (below) is unimportant.\par

Multi-band monitoring, including $g'$-band, was carried out over 5 nights with the LT \citep{mchardy_2023} and over 2 nights on the Faulkes North Telescope (FTN) \citep{montano_2022}. Both the LT and FTN are 2m telescopes. Given also that HiPERCAM has a higher throughput than the instruments on the LT and FTN, and no dead time, the HiPERCAM light curves are of much higher S/N so we do not consider the short timescale LT and FTN lightcurves further.

\subsection{Medium Timescales: days-month}

\underline{\bf TESS:~~}
A detailed light curve with 15min sampling for a month from the Transiting Exoplanet Survey Satellite (TESS) is shown  by \cite{burke_2020} and is the basis for their time series analysis.  TESS observes in a wide, 600-1000\,nm bandpass.
Variability amplitude and hence power spectral normalisation are a function of wavelength. In the standard 'rms/mean' power spectral normalisation which we use below we should therefore adjust the mean level of the TESS lightcurve by subtracting an appropriate constant so that the resulting power spectral normalisation is similar to that expected of a g'-band light curve covering the same frequency range. 

\begin{figure}
    \includegraphics[width=\columnwidth]{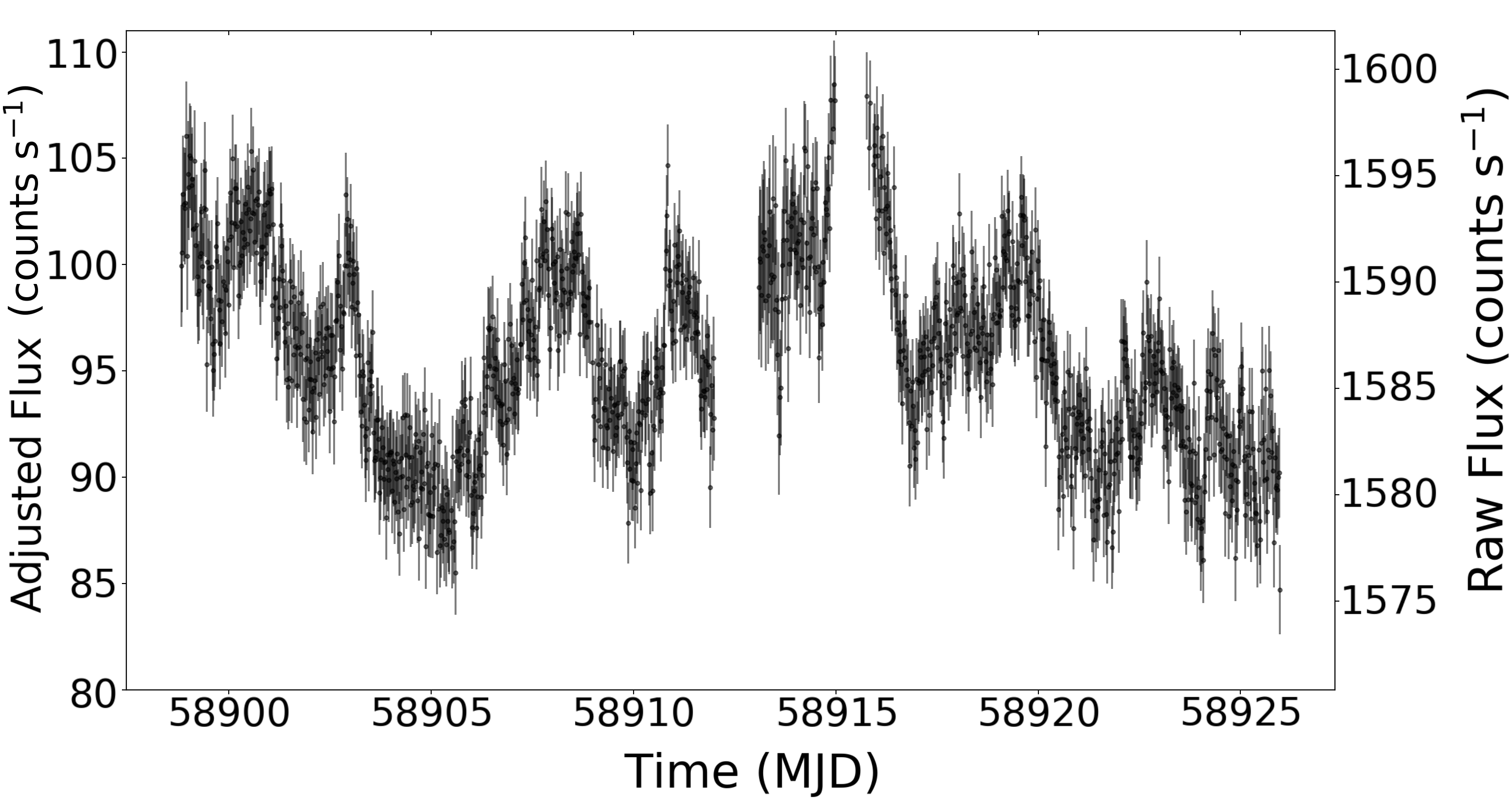}
    \caption{TESS light curve for NGC 4395 using data from \protect\cite{burke_2020}.
    The scale on the right-hand y-axis gives the original count rate. The scale on the left-hand y-axis gives the count rate after subtraction of the constant required to
    produce a light-curve with the same fractional variation as the LT g'-band over the same time range.  
    }
    \label{fig:TESSLC}
\end{figure}

We could calculate that constant by synthesising a white light lightcurve from the appropriately weighted {\it ugriz} HiPERCAM lightcurves and calculating the resultant PSDs. However there is an additional consideration with TESS in that the pixels are very large and hence there is a large non-AGN constant contribution to the light curve. This constant must also be removed. We can, in fact, correct for both problems by altering the zero level of the TESS lightcurve so that the resultant fractional variation, $F_{\rm var}$, i.e. standard deviation divided by the mean, of the TESS lightcurve is the same as that of the long term g'-band lightcurve discussed below (Sec.~\ref{sec:longtermlc}) over a similar length of time as that of the TESS lightcurve. The resultant light curve can be seen in Fig.~\ref{fig:TESSLC}.

\subsection{Long Timescales: months-years}
\label{sec:longtermlc}

\underline{\bf LT, LCOGT, ZTF, Zowada:~~} The main new data presented here are long term SDSS $g'$ band  light curves from the LT's IO:O camera \citep{steele_2004}, the LCOGT McDonald Observatory's Sinistro and Faulkes Telescope North's (FTN) Spectral cameras \citep{brown_2013}, the Zowada Observatory \citep{carr_2022} and the ZTF \citep{masci_2018}. The details of the observations can be found in Table \ref{fig:ObsTable}. \par

The LT and LCOGT data were processed with aperture photometry using the HiPERCAM pipeline, as described in \cite{mchardy_2023}, using the same comparison star, Star\,1, as shown in Figs.\,1 and 12 of \cite{mchardy_2023} and with the same source aperture radius of 4.25 arcsec. 
The ZTF light curve was extracted from the images by the Automated Learning for the Rapid Classification of Events (ALeRCE) pipeline \citep{masci_2018}\footnote{\url{https://alerce.science/alerce-pipeline/}} and the light curves were then downloaded from the ALeRCE ZTF Explorer\footnote{\url{https://alerce.science/services/ztf-explorer/}}. Zowada fluxes were also 
obtained by differential aperture photometry using in-house code which compares the AGN with a number of stars in the field \cite{carr_2022}. 

The light curves were initially in a variety of different units so to combine them we first converted them all into mJy. The LT and LCOGT light curves were initially in units of counts/sec relative to comparison Star\,1 and magnitudes are available for Star\,1 from SDSS which we converted to mJy. 
The ZTF light curve is in AB magnitudes which we again convert to mJy. The Zowada data were provided directly in mJy. 

\begin{figure}
    \includegraphics[width=\columnwidth]{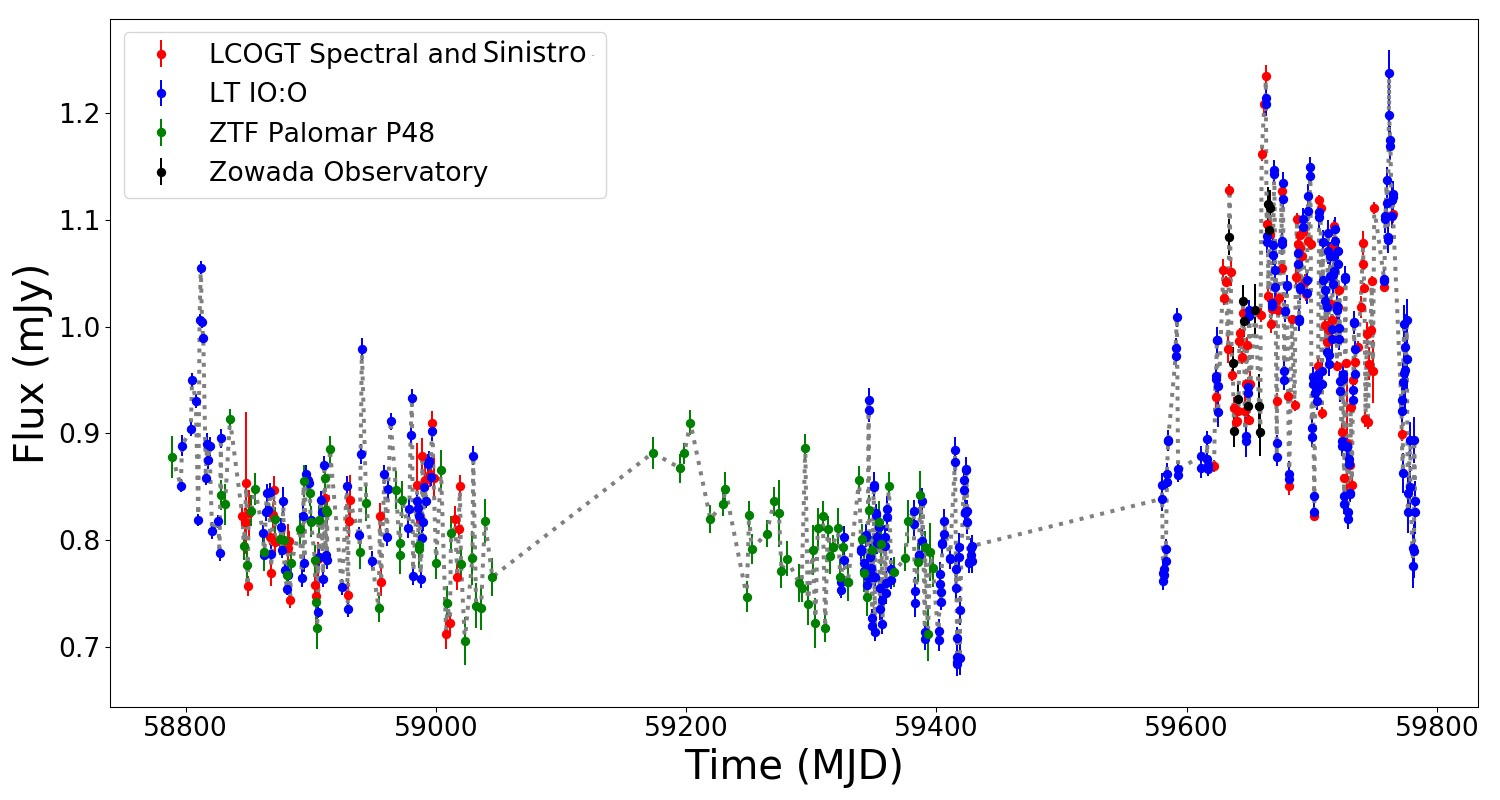}
    \caption{Long term SDSS g' band light curve for NGC 4395 comprised of data from Liverpool Telescope, Las Cumbres Observatory, Zowada Observatory, and the Zwicky Transient Facility.}
    \label{fig:LongTermLC}
\end{figure}

To ensure good cross-calibration between the various light curves, points with common time stamps were compared and then a Chi-squared minimisation of an additive and multiplicative translation was applied (relative to the LT data). This procedure created the light curve seen in Fig~\ref{fig:LongTermLC}. Due to seasonal breaks, the light curves are divided into 3 'epochs' by which we will refer to them.

\onecolumn

\begin{table}
\centering
\begin{tabular}{|c|c|c|c|c|}
\hline
\rowcolor[HTML]{9B9B9B} 
\textbf{Observation Name} &
  \textbf{Exposure Time} &
  \textbf{Number of Obs} &
  \textbf{Start Date} &
  \textbf{Stop Date} \\ \hline
LT IO:O & 40s (until 02/08/2021) / 60s (after) & 342 & 09/11/2019 & 22/07/2022 \\ \hline
LCOGT   & 45s (Spectral) / 240s (Sinistro)     & 128 & 28/12/2019 & 11/07/2022 \\ \hline
ZTF     & 30s                                  & 97  & 02/11/2019 & 02/07/2021 \\ \hline
Zowada  & $5\times 300$s                  & 13  & 24/02/2022 & 28/03/2022 \\ \hline
\end{tabular}
\caption{SDSS g' band Observations.}
\label{fig:ObsTable}
\end{table}

\twocolumn

\begin{figure}
    \includegraphics[width=\columnwidth]{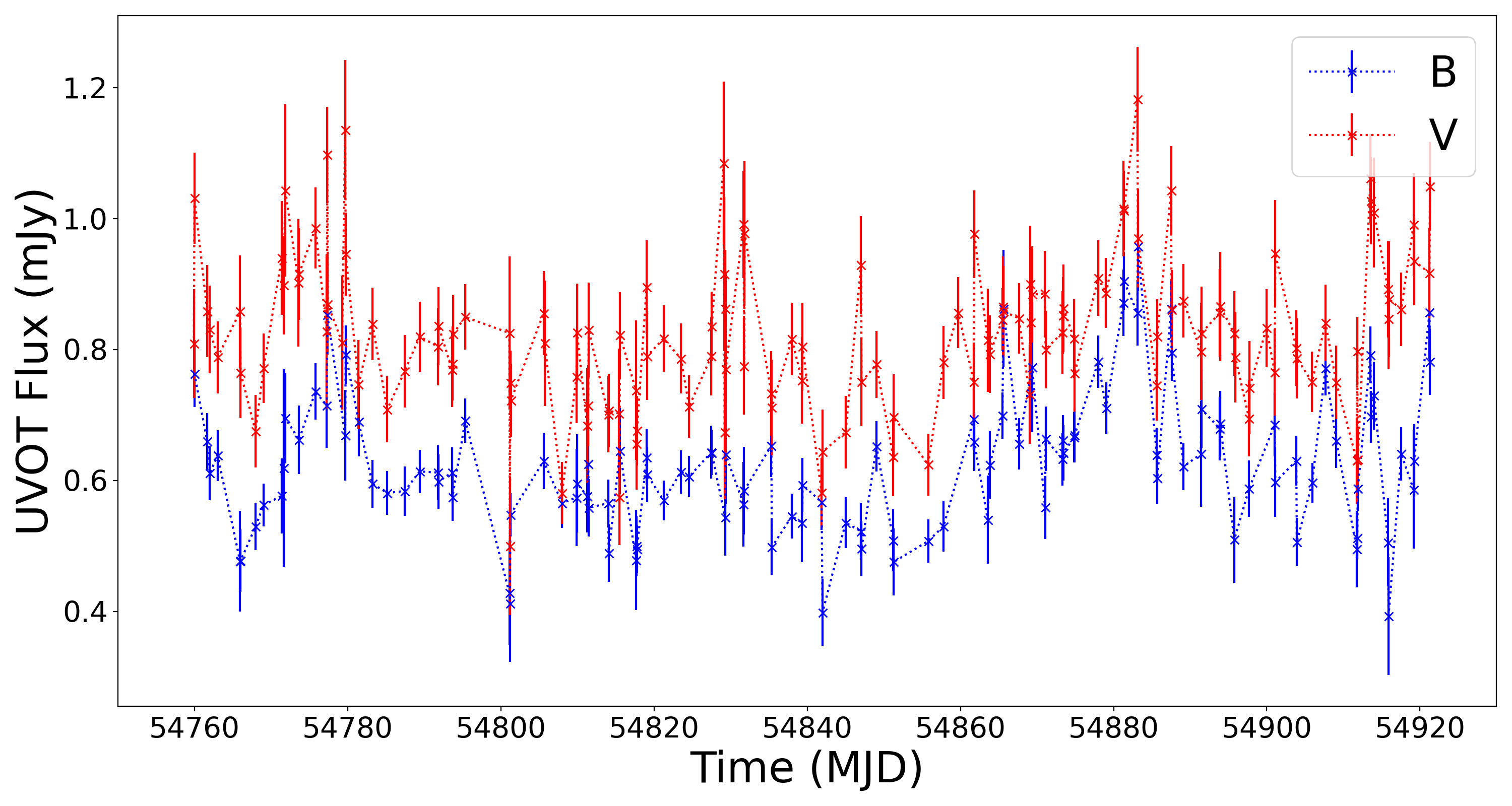}
    \caption{\swift UVOT light curve for NGC 4395 from \protect\cite{cameron_2012_correlated}.}
    \label{fig:CameronLC}
\end{figure}

\noindent
\underline{\bf Swift:~~}
There was long-timescale optical monitoring by Swift in 2008-2009 \cite{cameron_2012_correlated} over two periods of duration 112 and 161d, separated by a 76d gap, with typical 2d sampling. In Fig.~\ref{fig:CameronLC} we present again the Swift UVOT $B$ and $V$ band light curves from the longer, second, period.
The first period contains an outburst of amplitude larger than that seen in other observations of this source. The concern then is that the assumption made by most authors who analyse AGN temporal variability, including ourselves, i.e. that the variability is a statistically stationary process, might not be valid. However in Section~\ref{Analysis}, we show that the PSD of the first period is consistent with the PSD of the second period and so the outburst is probably just part of the normal variability of the source. Thus, although we note the caveat, we proceed on the assumption that the variability is statistically stationary and so normal PSD analysis methods may be used. The first period is, however, short and adds very little to our long-timescale temporal coverage so we do not in fact consider it elsewhere.

\section{Simulating Long-Term Optical Light Curves}
\label{method}

We simulate long-timescale optical light curves by reprocessing long-timescale X-ray light curves by an accretion disc model. 
The longest observed high quality X-ray light curves, from \xmmnogap, are only of maximum duration $\sim140$ks. Therefore we first have to simulate much longer X-ray light curves, of similar quality, but which include the long term variability properties.

\subsection{Simulating Long-Term X-ray Light Curves}

\subsubsection{Quantifying X-ray Variability}

The main variability characteristics of a light curve are usually quantified in terms of its power spectral density (PSD). AGN X-ray PSDs are often well described by bending power laws \citep{mchardy_2004}  given by
\begin{eqnarray}
    P(\nu) = A\nu^{-\alpha_{L}} \left(1 + \left(\frac{\nu}{\nu_{B}}\right)^{\alpha_{H} - \alpha_{L}}\right)^{-1} + C
    \label{singlebendPL}
\end{eqnarray}
where $\nu_{B}$ is the bend frequency, $\alpha_{H}$ and $\alpha_{L}$ are the high-frequency and low-frequency slopes respectively and $A$ and $C$ are constants.  As noted in \cite{mchardy_2004}, an arbitrarily large number of bends can be incorporated by daisy-chaining the above equation. 
A double-bending power law, which will be used later in this paper, is therefore given by
\begin{eqnarray}
    P(\nu) = A\nu^{-\alpha_{L}} \gamma^{-1} + C
    \label{doublebendPL}
\end{eqnarray}
with
\begin{eqnarray*}
        \gamma = \left[\left(1 + \left(\frac{\nu}{\nu_{B1}}\right)^{\alpha_{M} - \alpha_{L}}\right)\left(1 + \left(\frac{\nu}{\nu_{B2}}\right)^{\alpha_{H} - \alpha_{M}}\right)\right]
\end{eqnarray*}

If the PSD parameters are known, light curves similar to the observed light curves can be simulated using the method of \cite{timmer_1995}. However this method produces only Gaussianly-distributed light curves and the X-ray light curves NGC~4395 are not Gaussianly-distributed (i.e., to first approximation, they are not evenly distributed about the mean and are more `burst-like'). We therefore use the method of \cite{emmanoulopoulos_2013} where the parameters of the flux probability density function (PDF), as well as the PSD parameters, are included. 

These PSD and PDF parameters are derived directly from observations, here using our recent long \xmm light curves, using Python code\footnote{\url{https://github.com/samconnolly/DELightcurveSimulation}} written by \cite{connolly_2015}.
However these X-ray observations do not cover the longer timescales ($\sim$years) that we require for comparison with our observed long-timescale g'-band lightcurve. In particular the slope of the X-ray PSD on long timescales, $\alpha_{L}$, is not well determined. However we can estimate $\alpha_{L}$ by using the earlier Swift observations \citep{cameron_2012_correlated}.

\subsubsection{Measuring the low-frequency X-ray PSD slope}

The 4 new observed \xmm observations are well fitted (fit probability, P = 0.79) by a bending power law model with $\alpha_{L}= 0.8^{+0.2}_{-0.3}$, $\alpha_{H}=2.3^{+0.1}_{-0.2}$ and $\nu_{B}= 7.7^{+10.2}_{-6.5} \times 10^{-4}$\,Hz. These values are similar to those of \cite{vaughan_2005} for the same bending power law model. In particular \cite{vaughan_2005} give $\alpha_{L}= 1.20^{+0.40}_{-0.76}$ so their value and ours are similar to that of all other AGN where that slope is reasonably well measured. 

\begin{figure}
    \hspace*{-10mm}
    \includegraphics[width=\columnwidth,angle=270]{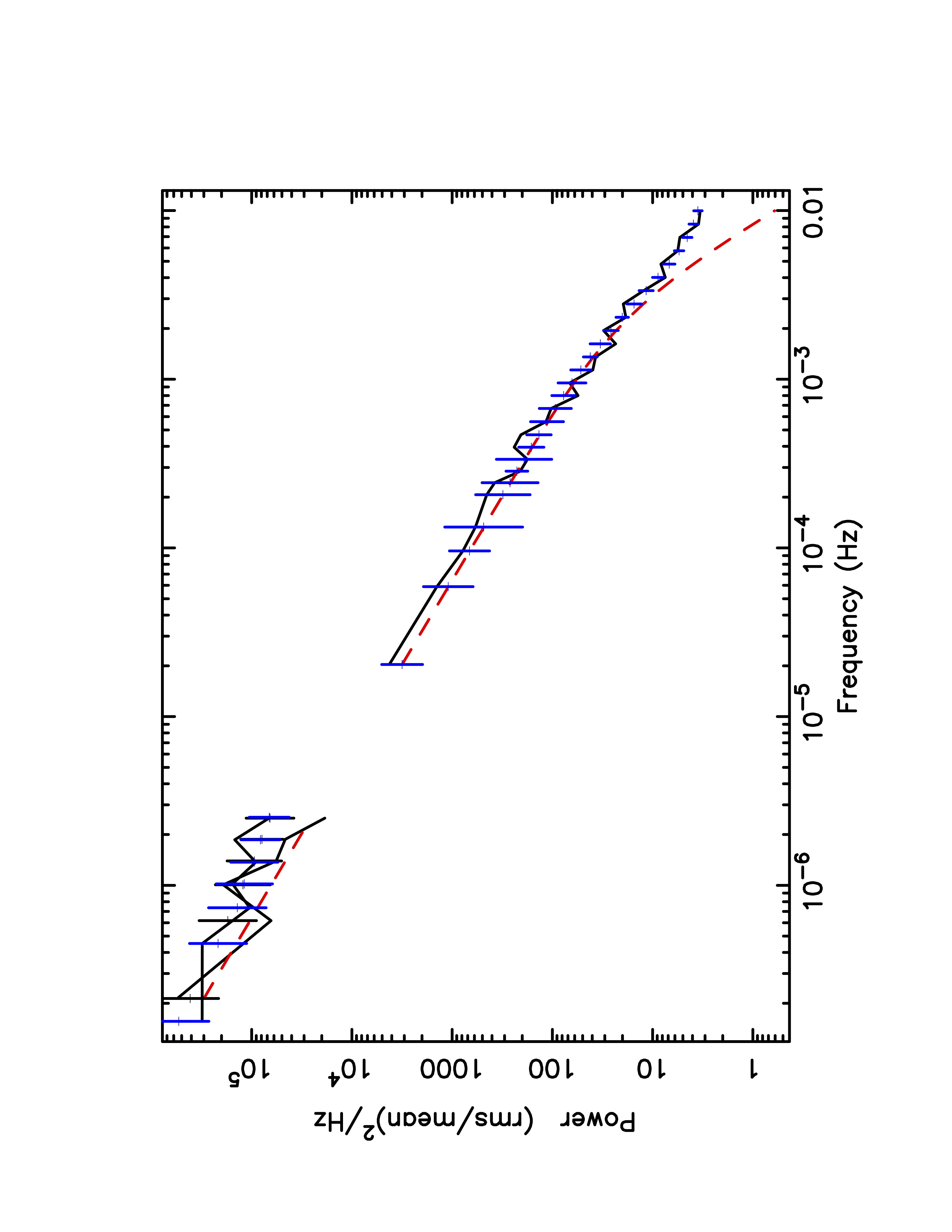}
    \caption{Combined Swift and XMM PSD.  The data are represented by the continuous solid black lines. The dirty model is represented by the errorbars and the underlying PSD is given by the dashed red line.}
    \label{fig:swift_xmm_psd}
\end{figure}

Although the sampling of the Swift data (2d) is much lower than that of the \xmm data, we can derive a combined Swift and \xmm PSD, which extends our measurement of $\alpha_{L}$ 
to years timescales using {\sc PSRESP}\footnote{\url{https://github.com/wegenmat-privat/psresp}}. 
{\sc PSRESP} properly takes account of red noise leak, aliasing and 
window effects in the PSD caused by the sampling pattern to reveal the true underlying PSD shape. We refer readers to \cite{uttley_2002}, and to other papers which have used {\sc PSRESP} \cite[e.g.][]{mchardy_2004}, for details of the method. The resulting PSD is shown in 
Fig.~\ref{fig:swift_xmm_psd} and is well described by a bending power law model. 

The average flux in the first part of the Swift light curve is about four times higher than in the second part. However the pattern of variability as quantified by the power spectrum  in standard units of $(\frac{rms}{mean})^{2}/Hz$, where the mean is the local mean of the particular part, is very similar and so we include both parts here separately.
The resulting 90\% confidence limits on the low-frequency slope are 0.75 and 1.1.
which is consistent with a fit where the low-frequency slope determined just from the \xmm fit extrapolates reasonably well to lower frequencies. Thus, for the present, we use the PSD and PDF parameters determined from fits to the \xmm data, assuming that $\alpha_{L}$ does extrapolate smoothly to years timescales, to simulate long-timescale X-ray light curves. Better sampled Swift observations are currently under way to improve our determination of the low-frequency X-ray PSD. When complete they will be the subject of future work but they are beyond the scope of the current paper.

\subsubsection{Comparison of observed and simulated X-ray PSDs}

Using the PSD and PDF parameters determined from those observations, and assuming that the low-frequency PSD is an extension of that measured with \xmmnogap, we can now simulate X-ray light curves which are arbitrarily long and arbitrarily well sampled. 

\begin{figure}
    \includegraphics[width=\columnwidth]{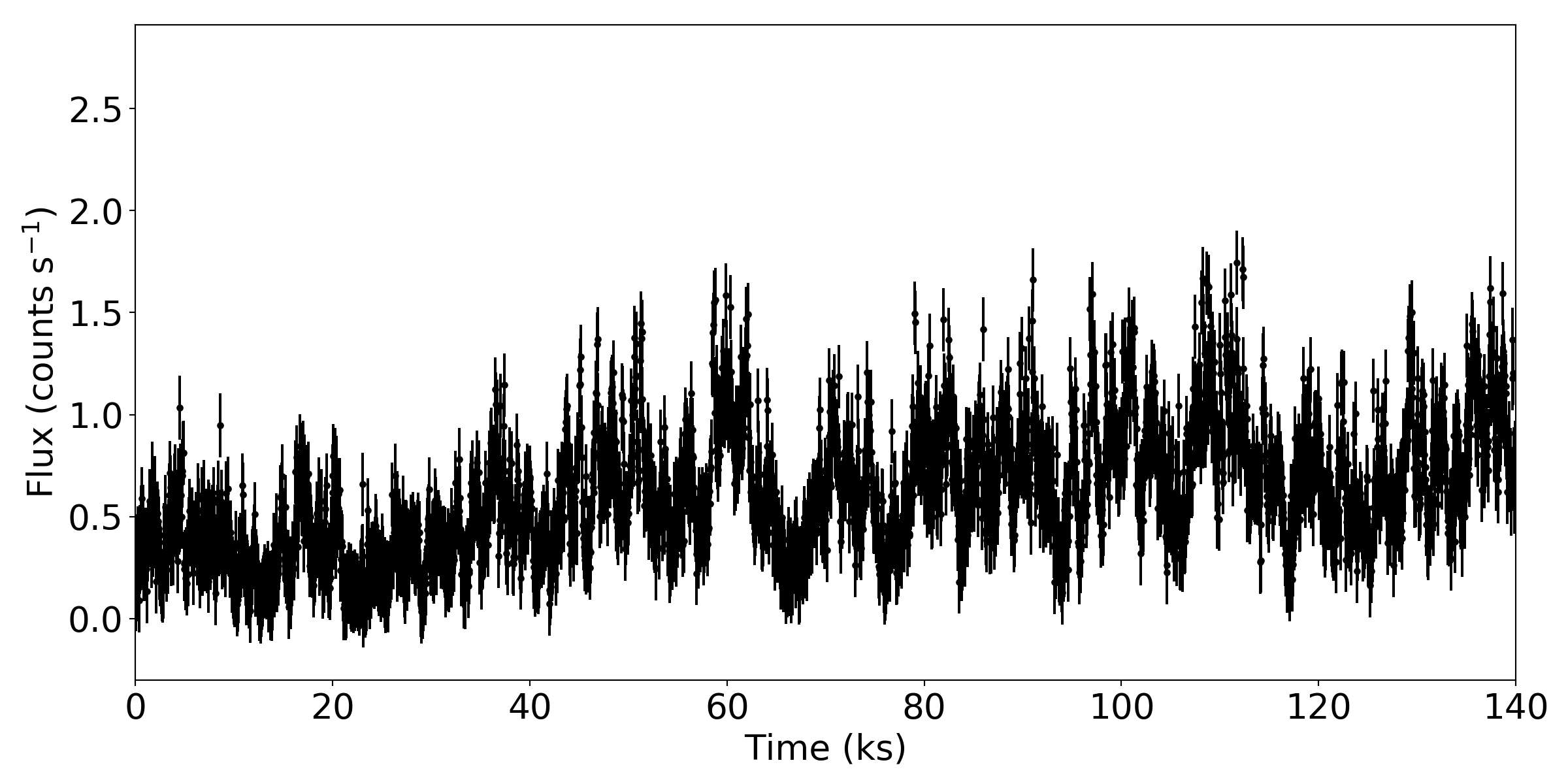}
    \caption{A 140ks sample of the simulated X-ray light curve using data from the night of January 2nd 2019, binned up to 50s for visual clarity.}
    \label{fig:XraySimLC}
\end{figure}

The main feature in the X-ray PSD is a bend at a timescale of $\sim1300$s. To simulate accurately this feature, including the PSD slopes both below and above this timescale, we retain the same 10s sampling as the original observed \xmm light curves. Although it is then quite computationally expensive, we simulate X-ray light curves of duration 10d which, following convolution with the g'-band disc response function (Sec.~\ref{sec:modelopt}), will produce optical light curves which contain many samples of all likely timescales of disc variability. An example of a simulated \xmm X-ray light curve of similar length (140ks) to the observed light curves is shown in Fig.~\ref{fig:XraySimLC}.

\begin{figure}
    \includegraphics[width=\columnwidth]{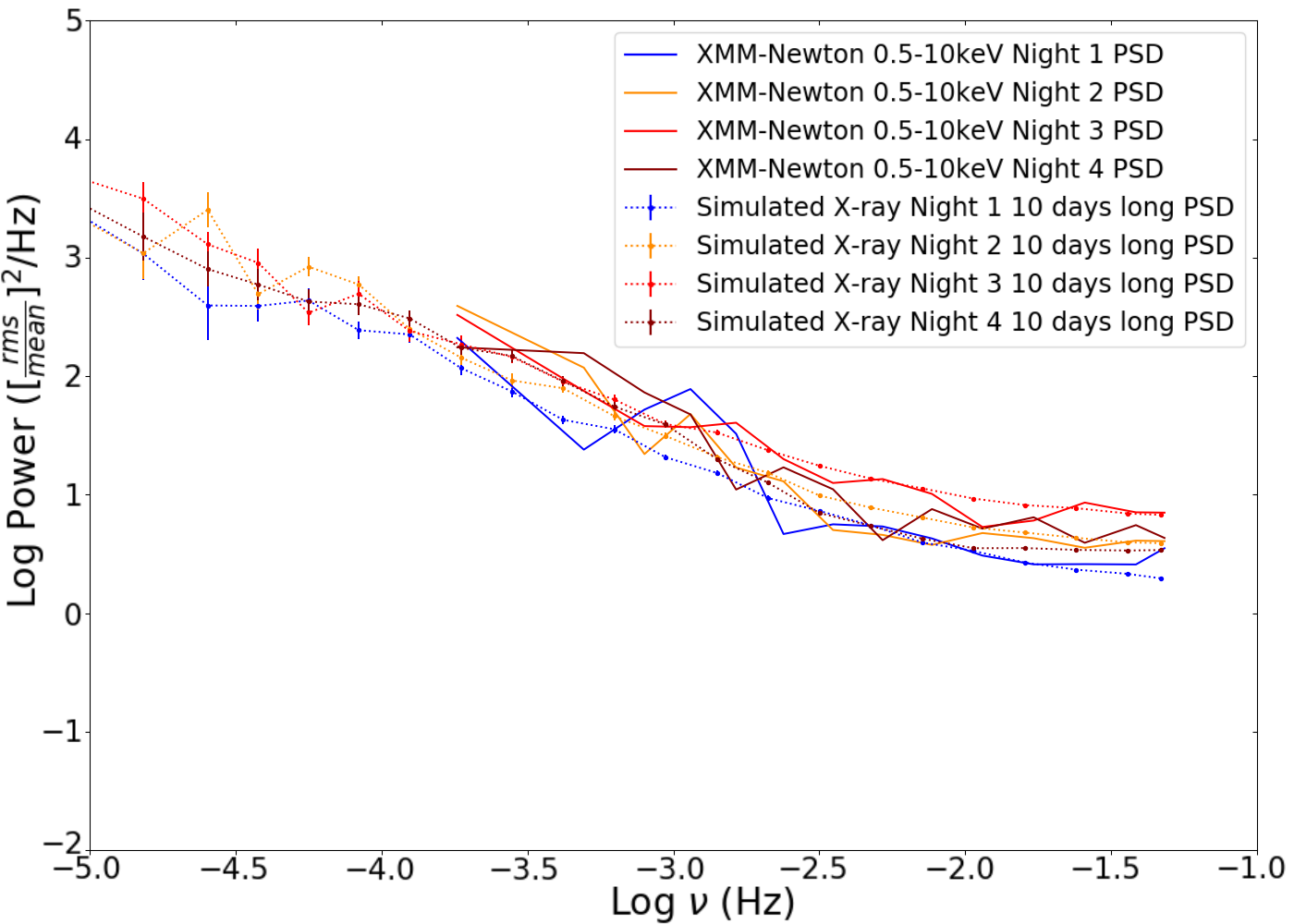}
    \caption{The 0.5-10keV X-ray power spectra from the \xmm data plotted with the PSDs of those same nights' simulated x-ray power spectra.}
    \label{fig:XrayPSD_DatavsSim}
\end{figure}

We simulate light curves based on each of the 4 separate \xmm observations to check for variation. 
To check that our analysis is self-consistent we show, in Fig.~\ref{fig:XrayPSD_DatavsSim}, the observed PSDs from our 4 \xmm observations and also the PSDs derived from 140ks sections of each of the simulated light curves. The observed PSDs all overlap and the simulated PSDs lie within the range covered by the scatter in the observed PSDs so, although not all identical, they are consistent with being part of the same variability process.

\subsection{Model Optical Light Curves from X-ray Simulations}
\label{sec:modelopt}

As the PSDs from all 4 \xmm observations are very similar then, to avoid unnecessary computation, we choose, somewhat arbitrarily, just one long X-ray lightcurve, simulated from the 2 January 2019 observation. We convolve this light curve with a model disc response function from KYNreverb \citep{kammoun_2021}. 
As our observed optical data is either $g'$-band or calibrated to be $g'$-band, we use a $g'$-band response function using the same input parameters as found in Beard et al. (in prep), i.e. zero spin, 10R$_{g}$ corona height, a colour correction factor, $f_{\textrm{col}}$, of 1.5, and a disc outer radius of 1500R$_{g}$. These numbers are very similar to those derived by \cite{mchardy_2023} from modelling the inter-band lags but are slightly modified to fit the new \xmm X-ray and UV lag data.  

The g'-band lags the X-rays by about 800s. \citep{mchardy_2016_the,mchardy_2023}. The observed lag corresponds to the median of the g'-band response function. \cite{kammoun_2021} show that the response from a disc drops, from the peak, to timescales $50 \times$ longer than the median lag (see their Fig.8), by a factor $\sim10^{4}$ which, for NGC4395, is $\sim 0.5$d. Thus our 10d long simulated light curves contain many samples of even this extremely low part of the response function and so are a good representation of what we might expect from reprocessing central X-rays by a surrounding accretion disc.
 
\begin{figure}
    \includegraphics[width=\columnwidth]{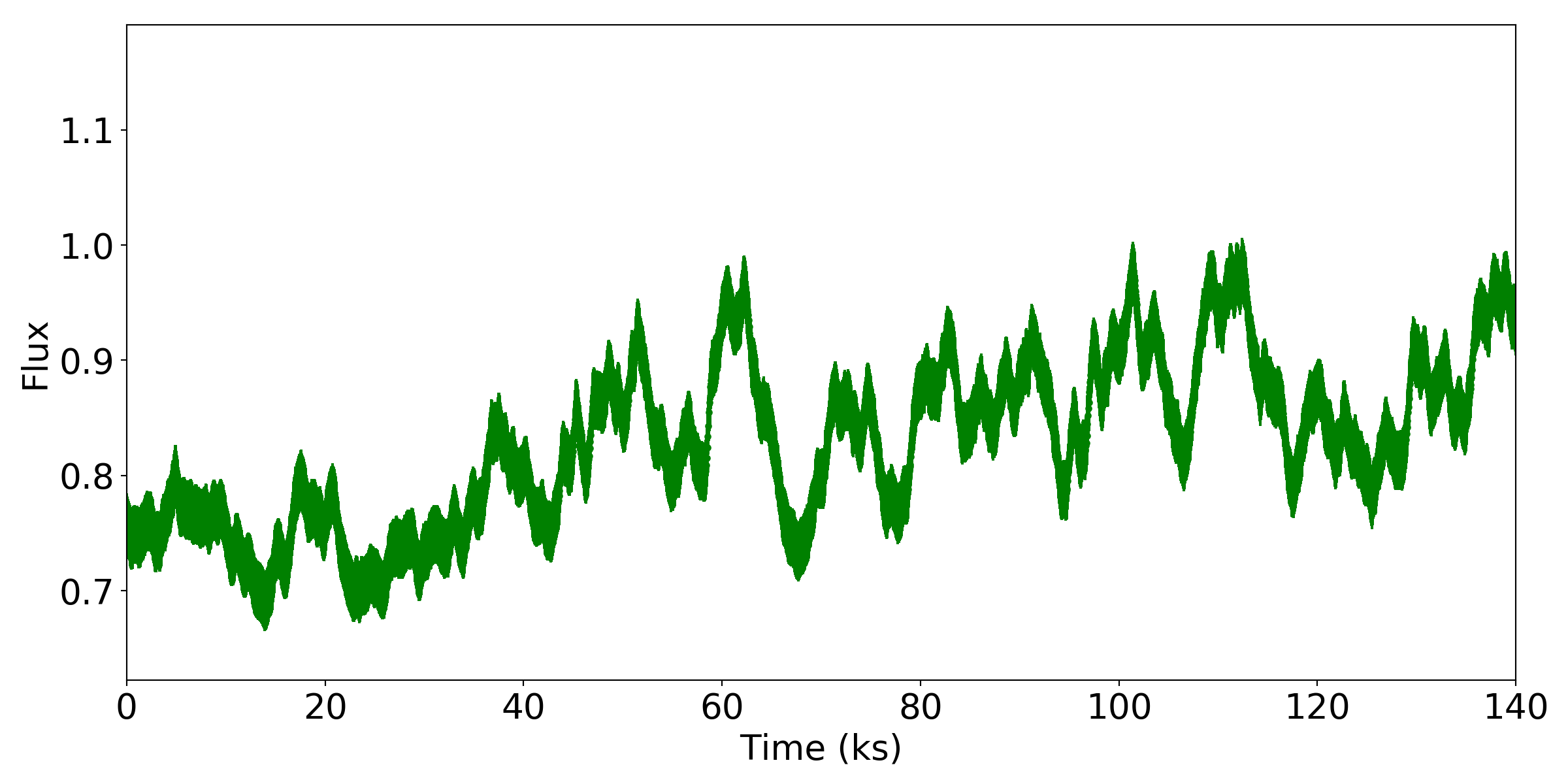}
    \caption{A 140ks sample of the synthetic $g'$-band light curve created from the simulated X-ray LC in Fig.~\ref{fig:XraySimLC}, binned up to 50s for visual clarity.}
    \label{fig:SyntheticGLC}
\end{figure}

For comparison with the simulated one-orbit \xmm  140ks X-ray light curve (Fig.~\ref{fig:XraySimLC}),
a 140ks sample of the synthetic $g'$-band light curve is shown in Fig.~\ref{fig:SyntheticGLC}.

\section{Comparison of Observed and Simulated Optical PSDs} \label{Analysis}

\subsection{Observed Optical Power Spectrum}

\subsubsection{The HiPERCAM PSD}
\begin{figure}
    \includegraphics[width=\columnwidth]{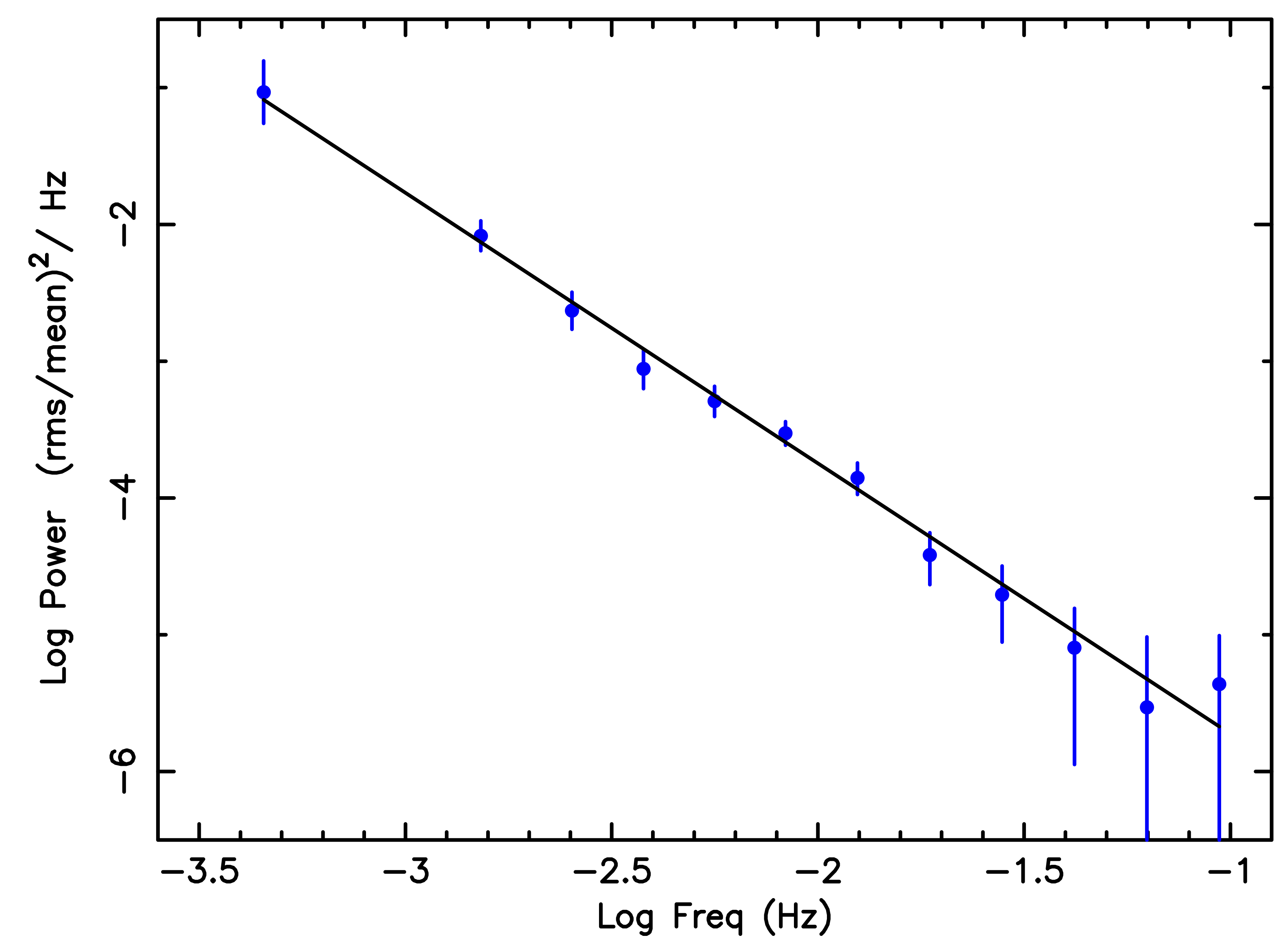}
    \caption{PSD of g'-band HiPERCAM observation with Poisson noise removed. }
    \label{fig:hipercam_psd}
\end{figure}

We begin by showing (Fig.~\ref{fig:hipercam_psd}) the PSD of the HiPERCAM $g'-$band data alone at the original 3s-sampling, produced using the standard method of \cite{deeming75} for discretely sampled data. This PSD is well fitted (P = 0.91) by a combination of a power law of slope $2.00\pm0.18$ together with Poisson noise. Here the Poisson noise power has been removed.  Aliasing does not affect these data but there may be a little red noise leak at the lowest frequencies. No complex modelling has been carried out.
For an underlying slope of 2, red noise leak does not change that slope, but it does increase the normalisation, which we will deal with below. 
The measured slope is in very good agreement with the value expected from a DRW (2.0), or the value of the high-frequency slope of the broken power law ($1.88\pm0.15$) fitted to the PSD of the TESS data \citep{burke21}, which covers the 2 decades of frequency immediately below the HiPERCAM range. Thus the initial indications are that a single unbending power law covers most of the frequencies sampled by TESS and HiPERCAM.

\subsubsection{{\sc PSRESP} modelling of the full dataset}
To measure the full optical PSD covering timescales from years to $\sim10$s, we combine the 
HiPERCAM and TESS data with other observations described in Section~\ref{obs} covering longer timescales.
To properly take account of PSD distortions caused by sampling irregularities, red noise leak, aliasing and Poisson noise from counting statistics which flatten the observed PSD at high frequencies, we model these data using {\sc PSRESP} \citep{uttley_2002}, following the prescription described by \citeauthor{uttley_2002} and used 
in previous papers \citep[e.g.][]{mchardy_2004}.
In {\sc PSRESP} the `dirty' PSD, derived from the 
raw observations and therefore distorted by the window function of those observations and including the contribution from Poisson noise, is first calculated. From an assumed underlying undistorted model PSD, light curves are simulated and sampled with the same window function as the observed data. The resulting simulated dirty PSD is then compared with the observed dirty PSD. The parameters of the underlying undistorted PSD are then varied to obtain the best match between the observed and simulated dirty PSDs. Following multiple simulations, uncertainties can then be derived on the simulated dirty model PSD, rather than on the observed dirty PSD. See \citeauthor{uttley_2002} for more details.

To cover the highest frequencies we include the HiPERCAM $g'$-band light curve of duration 8000s. As {\sc PSRESP} is computationally intensive we bin these data to 15s. 
For intermediate frequencies we input the TESS light curve, renormalised to $g'$-band variability, of duration 28 days, in $\sim$2ks bins. At the lowest frequencies we have the LT, LCOGT, Zowada and ZTF g'-band light curves which cover 3 years, but with seasonal gaps. 
We input the three seasonal sections as epochs 1, 2 and 3, binned at $\sim$1.75 days, which is the average sampling period. We also input separately the Swift B and V light curves from the second, longer, section. This section is of similar length to epochs 1, 2 and 3 so the binning is the same. Spectrally, the B and V bands are close to g'-band, on either side, so variability amplitudes should not be very different, with B expected to be slightly more variable and V less variable. As the number of data points is not large we do not renormalise them, with the post-facto justification that their resultant PSDs are almost identical.

For the very lowest frequencies longest timescales we input the entire 3 year  light curve in one piece. We bin it into 80d bins to cover the seasonal gaps and, with that large binning, we do not duplicate the frequency coverage obtained from the smaller binning of each individual epoch.
For the {\sc PSRESP} simulation of this light curve we set the sampling to a shorter timescale (0.3d) so that the high-frequency variations are included, and then bin up to 80d. This then gives us some useful extra long-term power in the log$(\nu / \rm{Hz}$)  range $-7$ to $-8$.

\begin{figure}
    \includegraphics[width=\columnwidth]{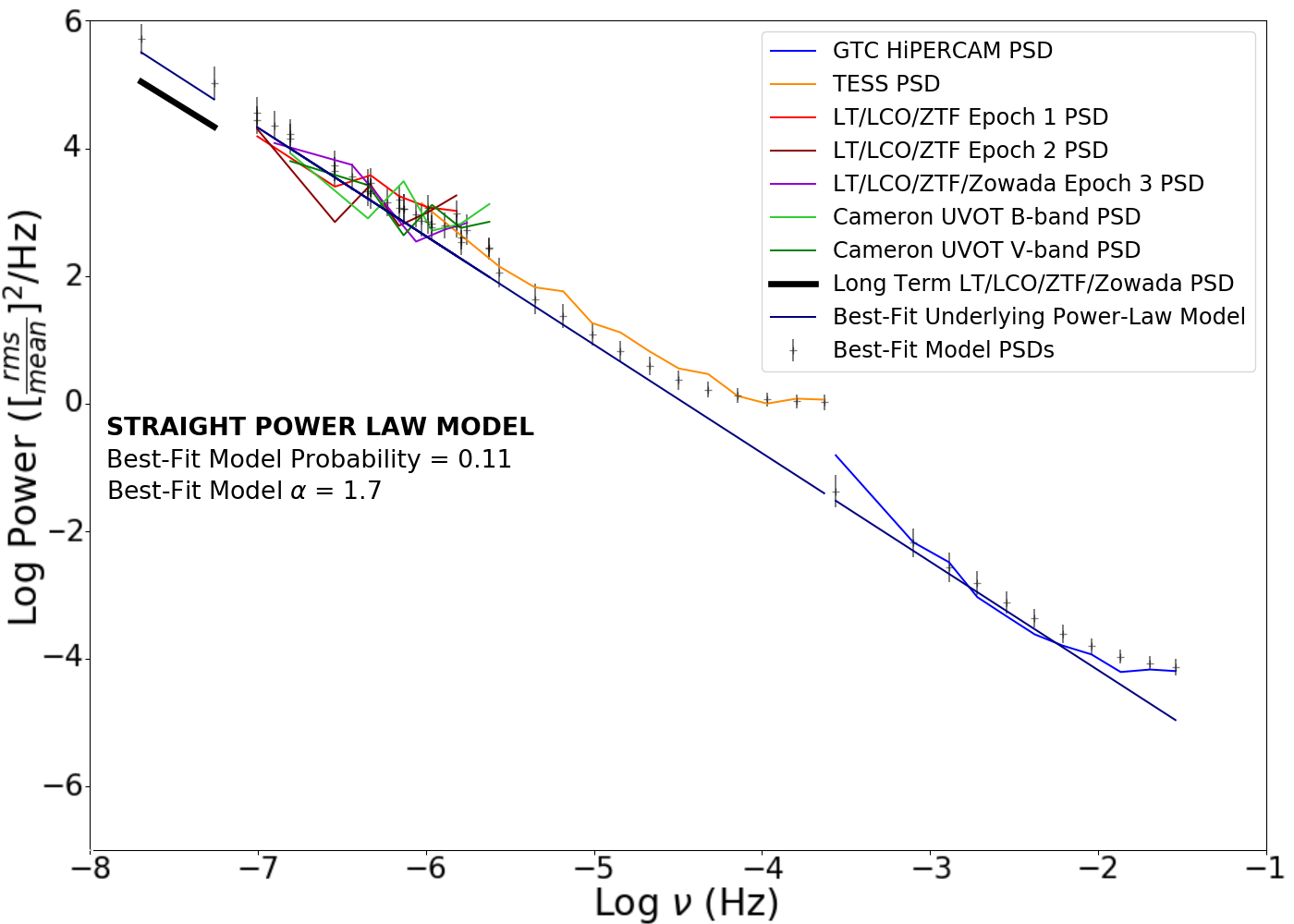}
    \caption{
    The observed 'dirty' optical power spectra are shown by continuous, zigzag, lines with different colours indicating the different source light curves as indicated in the legend. The corresponding best-fit model PSDs, assuming an underlying unbending power law PSD, are shown by the error bars. The underlying power law is shown by the smooth, continuous, straight blue line.  
    }    
    \label{fig:OptPSD_NoBend}
\end{figure}

\noindent 
\underline{\it Unbending Power Law Fit}\\
We first use {\sc PSRESP} to determine whether a simple unbending power law provides an adequate description of the data. The best-fit to such a model is shown in  Fig.~\ref{fig:OptPSD_NoBend}. The best-fit slope is $1.7^{+0.6}_{-1.0}$ with P = 0.11. Whilst not bad enough to entirely rule out this model, the fit probability is low and, visually, there appears to be a flattening at the lowest frequencies.

\noindent
\underline{\it Low Resolution Bending Power Law Fit}\\
We next fit the PSD with a bending power law.
Although the fit to an unbending power law had indicated that the deviation from a single power law occured towards the lower end of the frequency range which our data samples, we nonetheless began with a search for a bend over the full frequency range sampled from $10^{-8}$ to $10^{-1}$\,Hz. 
Performing a fit with high resolution in frequency over such a wide frequency range is computationally very time consuming so, to narrow down the parameter range to search over, we began with a search at low resolution.
We searched initially with resolution of a factor 2, i.e. 0.3 in log$(\nu)$, between searched frequencies. We sampled the high-frequency slope between 1.5 and 2.5 in steps of 0.5 and the low-frequency slope between 0 and 2.0 also in steps of 0.5. 
This search generated a best fit of P=0.57 with  
log$(\nu_{B}) = -5.6^{+2.4}_{-0.6}$
(i.e. $\nu_{B}=2.5_{-1.9}^{+640} \times10^{-6}$\,Hz),
$\alpha_{H}=2.0^{+0.5}_{-0.0}$ and $\alpha_{L}=1.0 \pm 0.5$. The large upper error on $\nu_{B}$ is a consequence of the large step sizes and poor resolution and the zero error on the lower limit to $\alpha_{H}$ is a result of hitting the end stops of the search range.

\begin{figure}
    \includegraphics[width=\columnwidth]{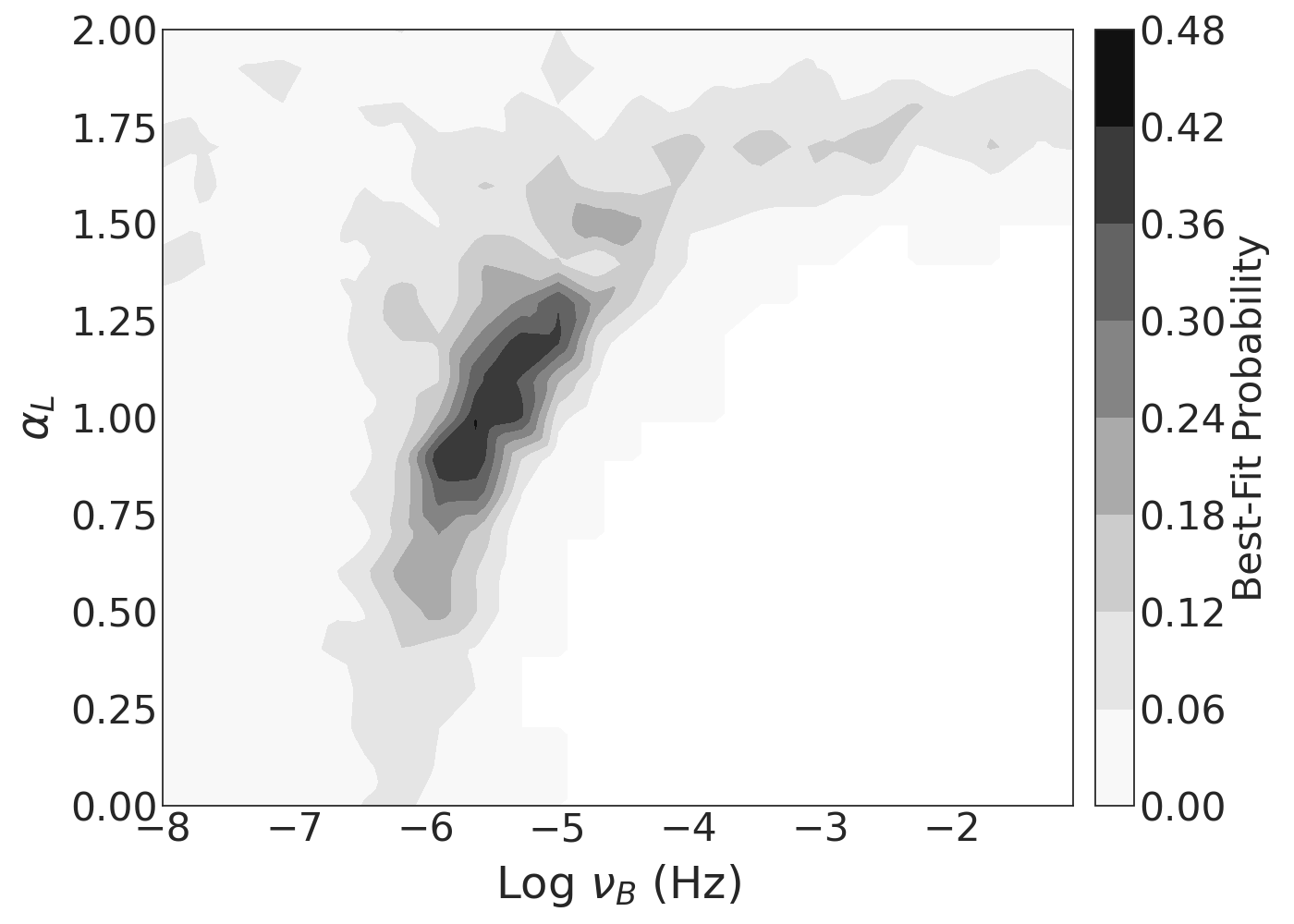}
    \caption{Best-fit probability contours of $\nu_{B}$ vs. $\alpha_{L}$ with $\alpha_{H}$ fixed at 2.0.  }
    \label{fig:OptPSD_fixedHF}
\end{figure}

\noindent
\underline{\it High Resolution Fit With Fixed $\alpha_{H}$}\\
As the best fit to $\alpha_{H}$  from both the HiPERCAM and TESS data individually is 2.0, and as the best fit value from our initial rough fit is also 2.0, we next fixed $\alpha_{H}$ at 2.0 and performed a search with higher resolution in  $\nu_{B}$ (factor 1.5) and $\alpha_{L}$ (steps of 0.1). A contour plot of the fit probabilities for these two parameters is shown in Fig.~\ref{fig:OptPSD_fixedHF}. This plot shows that $\alpha_{L}$ is very close to 1 and that $\nu_{B}$ lies in the range $10^{-7}$ to $10^{-5}$\,Hz.

\begin{figure}
    \includegraphics[width=\columnwidth]{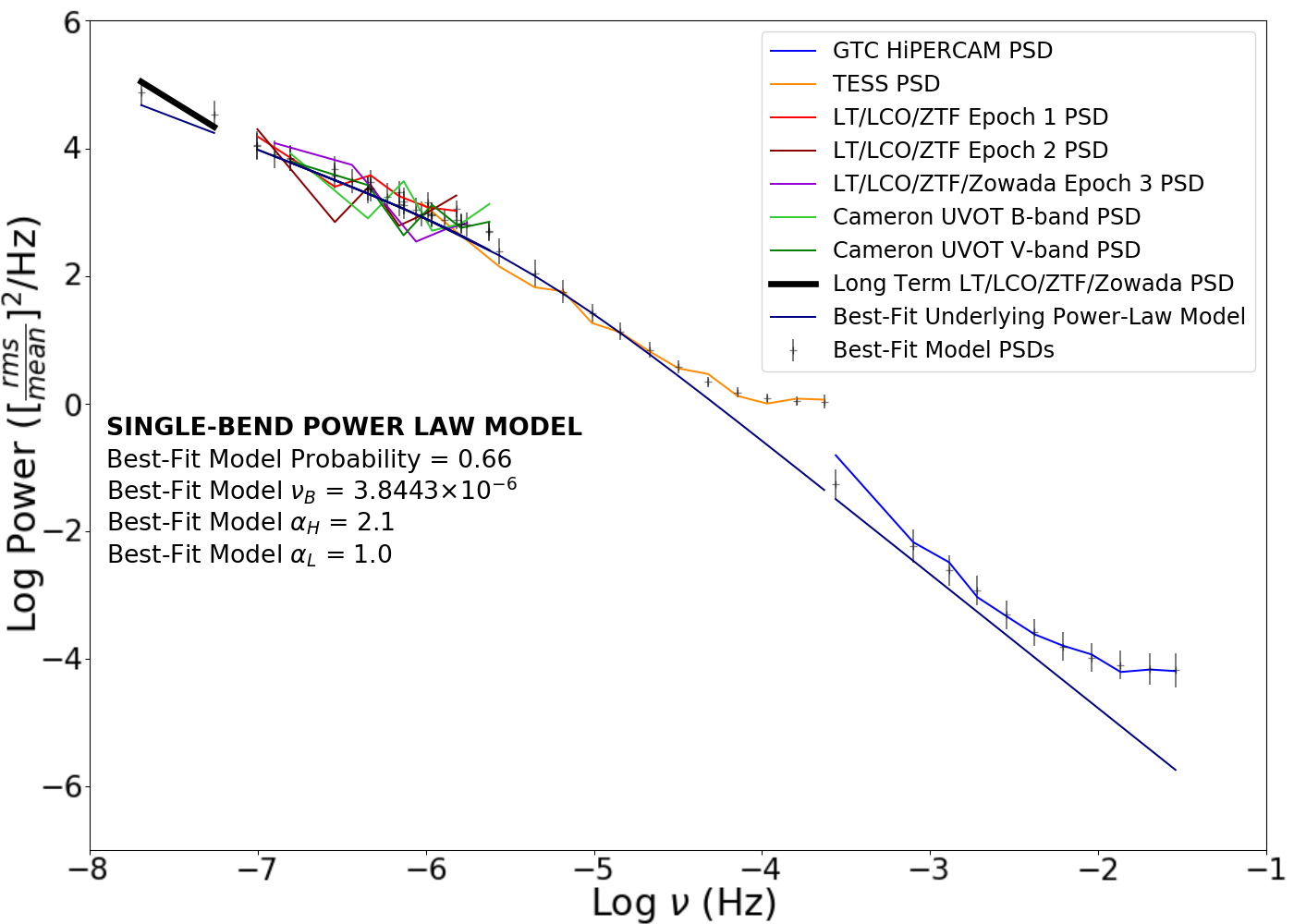}
    \caption{
    The observed 'dirty' optical power spectra are shown by continuous, zigzag, lines with different colours indicating the different source light curves as indicated in the legend. The corresponding best-fit model PSDs, assuming an underlying single-bend power law PSD, are shown by the error bars. The underlying power law is shown by the smooth, continuous, blue line with a single bend.
    }    

    \label{fig:OptPSD_SingleBend}
\end{figure}

\noindent
\underline{\it High Resolution Fit With All Parameters Free}\\
We finally performed a high resolution fit over all three parameters, restricting $\nu_{B}$ to the range $10^{-7}$ to $10^{-5}$\,Hz, with factor 1.5 steps in frequency. 
The search range for $\alpha_{H}$ is 1 to 4, in steps of 0.1 and the range for $\alpha_{L}$ is 0 to 2, also in steps of 0.1.
The resultant best-fit is plotted in Fig.~\ref{fig:OptPSD_SingleBend}.  The fit is very good, with a best-fit model probability of P = 0.66 with $\alpha_{H} = 2.1^{+0.2}_{-0.4}$,  $\alpha_{L} =1.0^{+0.2}_{-0.2}$ and 
$\nu_{B} = 3.8^{+4.8}_{-2.6}\times10^{-6}$\,Hz, corresponding to a bend timescale of $3.05^{+6.6}_{-1.7}$\,d.

Taking the bend timescale as $1/\nu_{B}$, we obtain a value of 3d, which is very similar to the 2.3d DRW damping timescale measured by \cite{burke_2020}. \citeauthor{burke_2020} also fit what they refer to as a broken power law which, functionally, is the same as the bending power law used here. They list the break frequency as $0.114 \pm0.066$ day$^{-1}$ which, taking the simple reciprocal as a measurement of the timescale, for comparison with our measurement, gives a bend timescale of 8.8d.
However \cite{burke_2020} do not make strong claims for the precision of their bend timescale as the bend is very close to the lower limit of the frequencies covered by the TESS data, which are better covered here.

The high-frequency slope measured here is in very good agreement with that ($1.88\pm 0.15$) measured by \cite{burke_2020} from their broken power law. The main difference between the present analysis and that of \cite{burke_2020} is that we find a low-frequency PSD slope of $1.0\pm0.2$ rather than the $0.00\pm0.86$ of the broken power law or the 0 required by the DRW model. However again we note that the TESS data provide very little PSD frequency
coverage below the bend, whereas here we have nearly 2 decades. Thus the uncertainty in the low-frequency PSD slope of \cite{burke_2020} is sufficiently large to allow compatibility with the present results.

A high-frequency slope of 2 is a fixed parameter of a DRW model but just because a DRW model can be fitted to a lightcurve, it does not mean that the high-frequency PSD slope is exactly 2. The most that we can say is that a measured slope of 2 is consistent with a DRW model.

As we now have a good fit to the observed optical PSD we proceed, in Section~\ref{sec:synthetic_psd}, to determine whether the model PSD derived from X-ray disc reprocessing is consistent with the observed PSD.

\subsection{Synthetic Optical Power Spectra}
\label{sec:synthetic_psd}

\begin{figure}
    \includegraphics[width=\columnwidth]{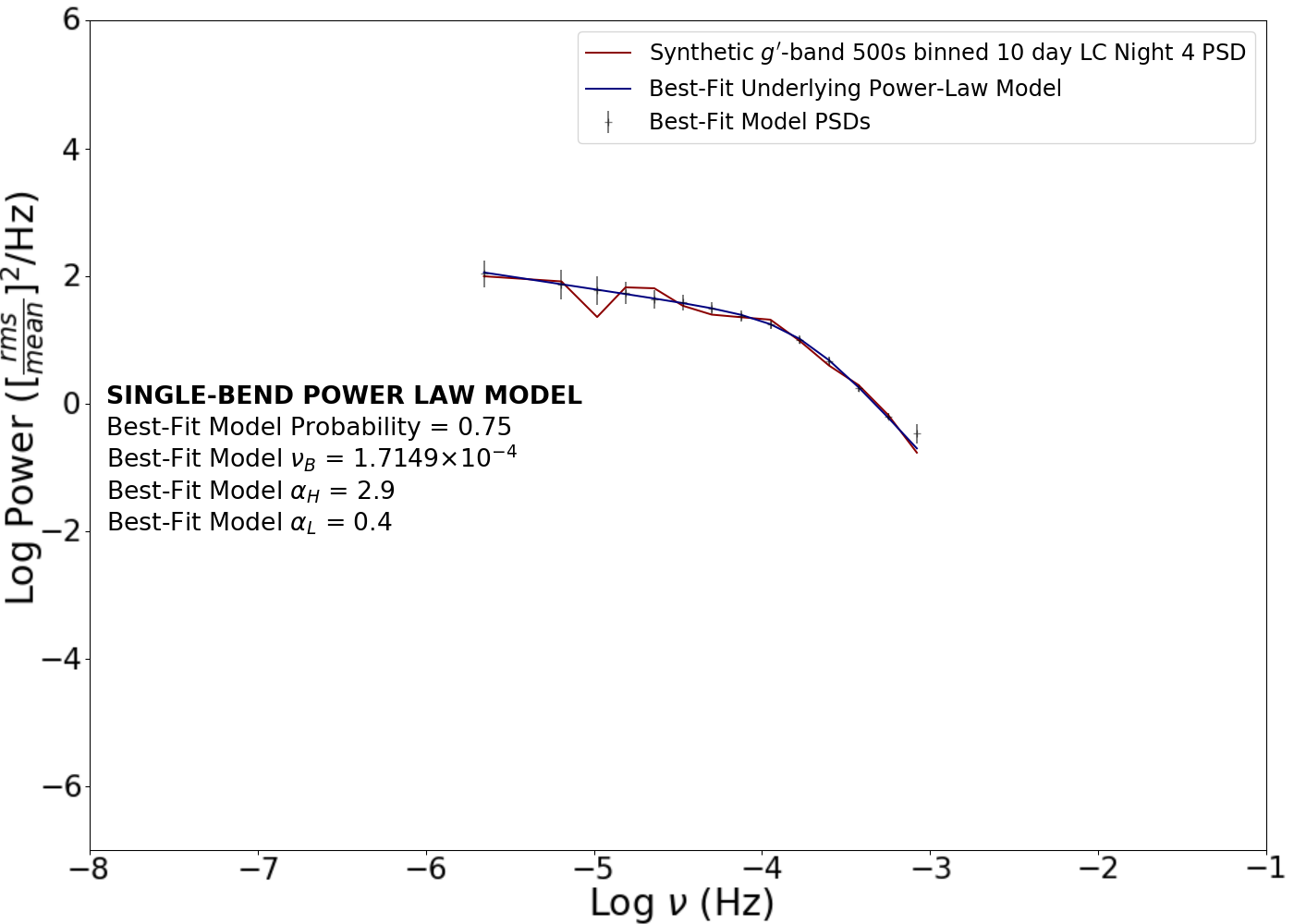}
    \caption{The synthetic $g'$-band PSD from the synthetic 10d lightcurve based on the \xmm night 4 data. Also plotted are the best-fit model PSD as well as the best-fit underlying power law for a bending power law model. This plot uses the same axes limits as the full observed optical data PSD for ease of comparison.}
    \label{fig:SyntheticG_PSD}
\end{figure}

In Fig.~\ref{fig:SyntheticG_PSD} we present the PSD resulting from the synthetic optical 10d $g'$-band light curve, a section of which is shown in Fig.~\ref{fig:SyntheticGLC}, together with the best fitting model which has
$\nu_{B} = 1.72^{+0.36}_{-0.75} \times 10^{-4}$\,Hz (i.e. a timescale of 1.6\,hr), 
$\alpha_{H} = 2.9^{+0.1}_{-0.3}$ and
$\alpha_{L} = 0.4^{+0.1}_{-0.3}$. 
This PSD shows a clear bend at $\sim10^{-4}$\,Hz with a steep high-frequency slope and fairly flat low-frequency slope. This PSD looks similar to the X-ray PSD from which it was generated except that the bend is a decade lower in frequency. The high-frequency slope is also slightly steeper consistent with the expectations from the disc reprocessing. The very pronounced flattening at $\sim10^{-4}$\,Hz in the synthetic PSD is not obviously visible in the observed optical PSD. 

The optical lightcurves from which this PSD were derived were simulated from the X-ray lightcurves assuming a reprocessing function based on a mass of $3.6\times 10^{5}$\msun and a low accretion rate. However if the mass is actually a decade lower, it will make little difference to the optical lightcurves and hence to synthetic PSD as the disc reprocessing function is mainly defined by the lags between wavebands, which are fixed. Thus parameters such as the accretion rate and possibly the colour correction factor \citep[see][]{mchardy_2023} would have to be adjusted to give the same lags.

One possibility is that this bend does exist in the observed PSD but that, in addition to disc reprocessing, there are other causes of variability which produce longer term variations, leading to the bend we detect at $\sim$10$^{-6}$\,Hz and obscuring the bend at $\sim$10$^{-4}$\,Hz. In the next section we explore whether both bends could be present in our data.

\subsection{Testing for double-bend compatibility}
\label{sec:testing_double-bend}

\begin{figure}
    \includegraphics[width=\columnwidth]{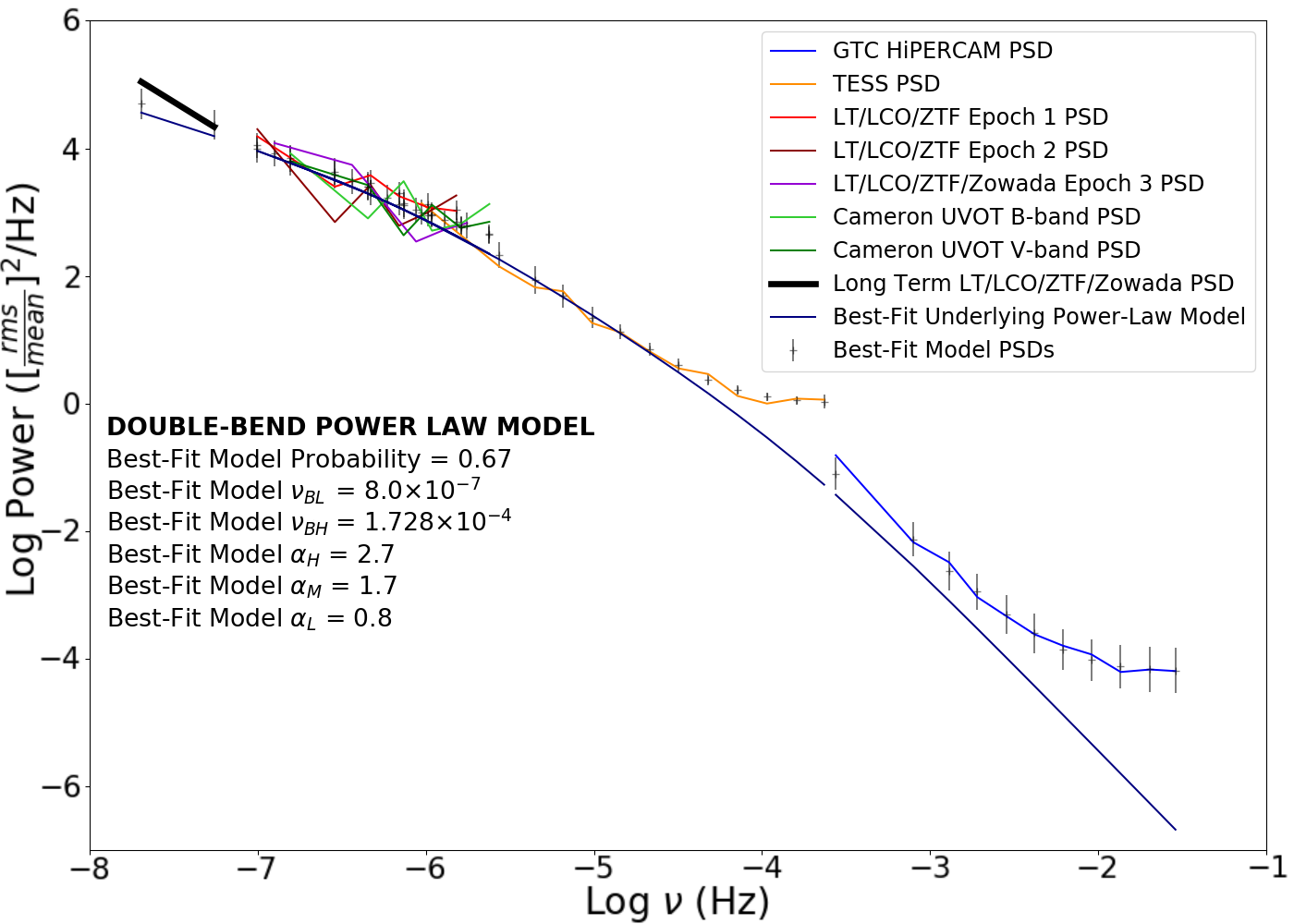}
    \caption{The optical power spectra from our data plotted with their best-fit model PSDs as well as the best-fit underlying power law for a double-bending power law model.}
    \label{fig:OptPSD_DoubleBend}
\end{figure}

In Fig.~\ref{fig:OptPSD_DoubleBend} we show the best fit for a PSD with two bends. All parameters were allowed to vary.
The low-frequency bend is at $\nu_{BL}= 8.0^{+2.4}_{-4.3} \times 10^{-7}$\,Hz and the high at $\nu_{BH}=1.7^{+0.4}_{-0.6} \times 10^{-4}$. The slope below the lowest bend, $\alpha_{L}= 0.8^{+0.3}_{-0.3}$, the slope between the two bends, $\alpha_{M}=1.7^{+0.3}_{-0.2}$, and the slope above the highest bend, $\alpha_{H}=2.7^{+0.3}_{-0.5}$. 
The fit probability (P = 0.67) is almost identical to the fit probability for a single bending power law so it is not, statistically, a better fit. It is, however, still a good fit and the inclusion of a second bend at a frequency near to that expected from reprocessing is consistent with the data.
Thus a combination of reprocessing to explain the high-frequency optical variability and some other variability process, e.g. intrinsic disc fluctuations, could explain the total observed optical variability.

\subsection{The origin of the low-frequency optical variations}

To confirm that a combination of reprocessing and a second independent source of variations can explain the observed variability, and to roughly estimate the relative contributions of these two processes, we attempt to reproduce the observed PSD with two such light curves. For the reprocessed light curve we use the 10d one discussed above in Sec.~\ref{sec:modelopt}. As there is no obvious sign in any of the observed optical lightcurves of the non-linear variations seen in the X-rays, we can simulate the independent source of variations extending to low frequencies using the method of \cite{timmer_1995}. We take  $\alpha_{L}=1$, $\nu_{B}= 1 \times 10^{-6}$\,Hz and $\alpha_{H}=2$. 

In the standard PSD normalisation of (rms/mean)$^{2}$/Hz which we use here, the power level of the PSD can be adjusted by changing either the rms or the mean of the input light curve, or both. Here, for the independent source of variations, we take the same mean level as for the reprocessed light curve and alter the rms until we obtain a combined PSD (Fig.~\ref{fig:SynthTKPSD_DoubleBend}) which is similar to the observed one.  We find that an rms of approximately 70 per cent of the mean, when combined with the reprocessed light curve, is required. This adjusted rms is about 10 times the rms of the reprocessed lightcurve. However the exact value should not be regarded as being particularly precise as there are a number of uncertainties involved, for example the exact shape of the reprocessing function which depends on unknowns such as the colour correction factor.

Although, for the computational reasons stated earlier, this PSD does not extend to quite as low frequencies as our observed PSD it does extend far enough for us to determine whether the two combined lightcurves might provide an explanation of the observed PSD, which it does. We do not claim that this combination is the only possible explanation, just that it is one possible explanation.

\begin{figure}
    \includegraphics[width=\columnwidth]{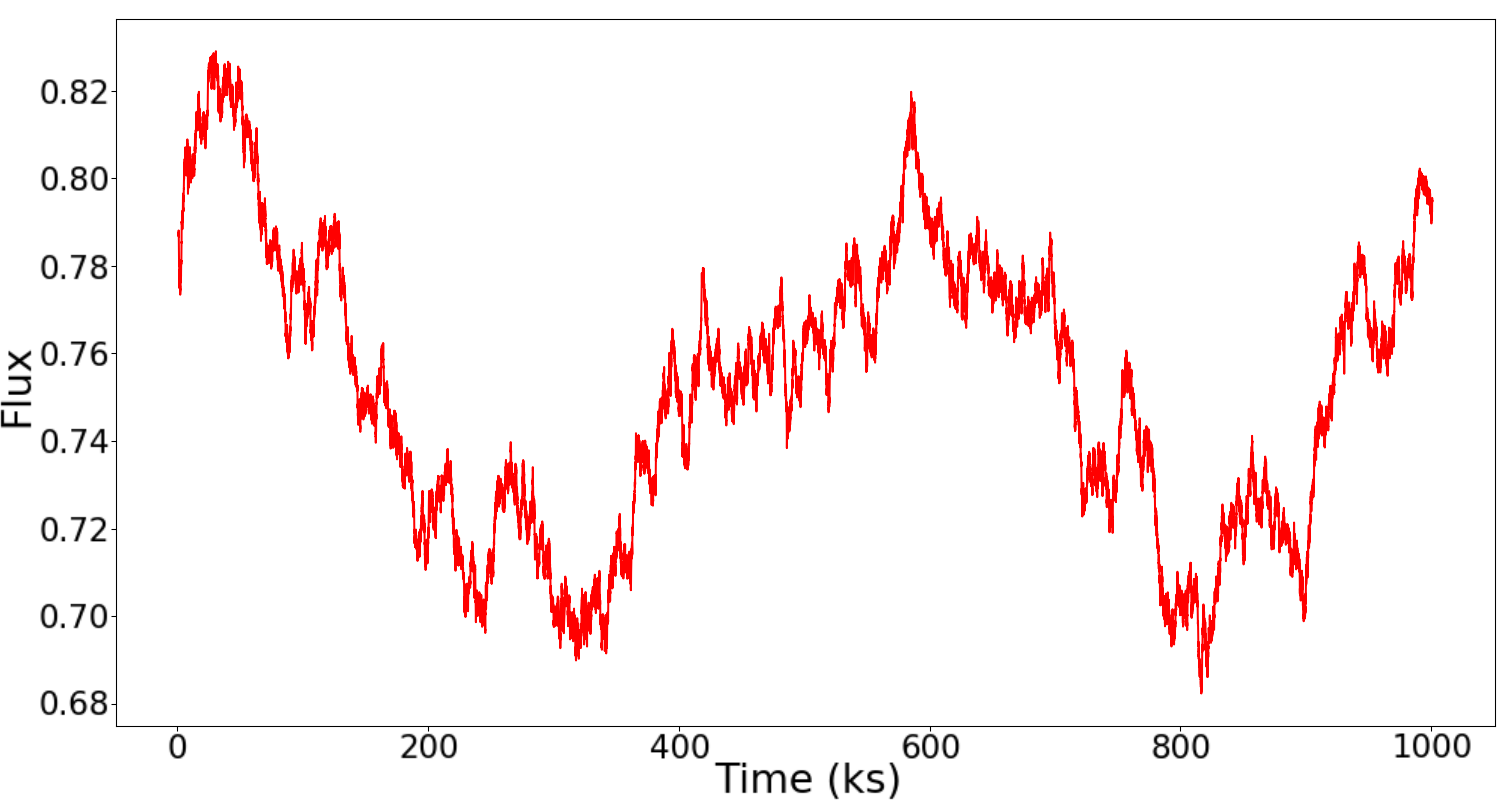}
    \caption{Simulated 10 day light curve using the \protect\cite{timmer_1995} method with an input bending power law with bend frequency $10^{-6}$, high-frequency slope -2, and low-frequency slope -1.}
    \label{fig:TKLC}
\end{figure}

\begin{figure}
    \includegraphics[width=\columnwidth]{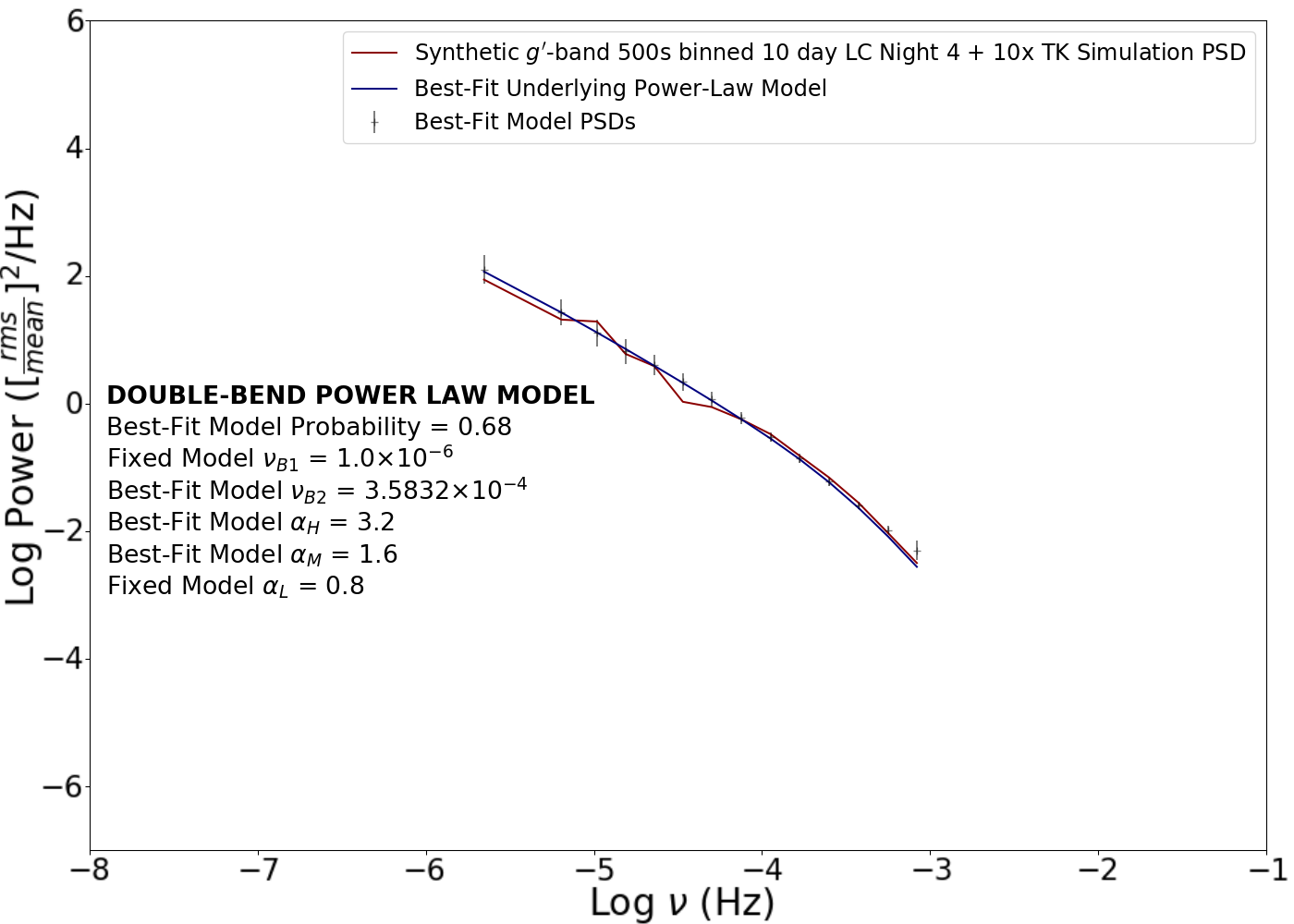}
    \caption{The power spectra from our synthetic $g'$-band light curve plus 10$\times$ the \protect\cite{timmer_1995} simulated light curve plotted with its best-fit model PSD as well as the best-fit underlying power law for a double-bending power law model. This plot uses the same axis limits as the optical data PSD for ease of comparison.}
    \label{fig:SynthTKPSD_DoubleBend}
\end{figure}

\subsection{Comparison with previously published Optical PSDs}
The optical PSD which we have derived here is of very high quality and is well described by a bending power law with $\alpha_{L} =1$, $\alpha_{H} =2$ and a bend timescale of 3d. We briefly compare this PSD with the previously published good quality optical PSDs noted in Sec.~\ref{intro} to see whether a similar model, and hence a similar underlying physical mechanism, might apply to them all. 

Considering first the bend timescales,
with our extended low-frequency data we confirm 
the timescale measured in NGC~4395 from TESS data alone using DRW modelling by \cite{burke_2020}. From the present data we find no requirement for any further bends at lower, or higher, frequencies. Although choice of black hole mass can alter the slope of the relationship a little, NGC~4395 fits into the relationship between AGN mass and bend (or damping) timescale which has already been thoroughly discussed by \cite{burke21} and so is not unusual as far as bend timescales go.

%\subsubsection{The high-frequency PSD slope}
The high-frequency slope in NGC~4395 is towards the shallow end of the distribution of Kepler high-frequency slopes \citep{smith_2018, mushotzky_2011}, but is within their distribution. The slope which we simulate based on X-ray reprocessing, 2.9, is towards the steeper end of the Kepler slope distributions but is, again, within the distribution. 
%Although only measured over 1.5 decades, \cite{panagiotou_2020} measured slopes of $\sim2$ in both optical, UV and X-ray bands in NGC~5548.
Thus reprocessing of X-rays is consistent with producing at least some of the high-frequency optical variability but is not necessarily the main contributor.

%\subsubsection{The low-frequency PSD slope}
Low-frequency optical PSD slopes are rarely measured because PSD bends tend to be near the lower end of the range of timescales sampled. NGC~4395 is a rare exception. The only large sample we are aware of where low-frequency slopes are listed is that of \cite{simm16}. The sampling of their Pan-STARRS lightcurves \citep{simm15}, typically 70-80 observations over 4yr though not very evenly spread, required use of CARMA models, rather than Fourier techniques, to estimate PSD parameters. They find an average $\alpha_{L} \sim 1$. \cite{smith_2018} show bending PSDs for 6 AGN but do not list low-frequency slope values. However four of the six AGN where bending power laws are required appear to have slopes which are rising towards the lowest frequencies (their Fig.10), although 2 are falling. Thus the limited previous data is consistent with the better measured $\alpha_{L}$ presented here.

The one AGN which does not easily fit the pattern discussed above, or is at least an outlier to the distributions, is Zw 229-15, which is probably the best-observed Kepler AGN. 
In the analyses of both \cite{edelson_2014} and \cite{smith_2018} its PSD shows a clear bend.
\cite{smith_2018} list a power law slope of 3.4, although it is not clear whether this is a fit to an unbending power law or to the high-frequency part of a bending power law. However \cite{edelson_2014} clearly list $\alpha_{H} \sim 4$. This slope is at the extreme limits of all listed high-frequency AGN optical PSD slopes. At low frequencies \cite{edelson_2014} list $\alpha_{L} \sim 2$, steeper than any previous low-frequency slopes. From Fig.10 of \cite{smith_2018}, their low-frequency slope appears somewhat flatter but they do not give a slope value.

A possible explanation of the extreme slopes in Zw 229-15 might be if its mass is actually much higher than the $\sim 10^{7}$\msun noted by  \cite{mushotzky_2011} and so 
the PSD bend detected by \cite{edelson_2014} and \cite{smith_2018} is actually the equivalent of a second, higher frequency bend which, in lower mass AGN like NGC~4395, would be at too high a frequency to be detected. The low-frequency slope observed in Zw 229-15 and the high-frequency slope in NGC~4395, both having the same slope value of 2, would then just be intermediate slopes in a twice bending power law model. However with the present mass ($\sim 10^{7}$\msun), and with the presently observed bend timescale, Zw 229-15 fits in well to the mass-timescale relationship of \cite{burke21}. There is also no sign of a second high-frequency bend in NGC~4395, even to 4 decades above the present bend. Thus it is not easy to argue for the second bend hypothesis. 

Measurement of PSD parameters is quite dependent on the exact model being fitted and the production of reliable long-timescale light curves from Kepler data is not trivial \citep[e.g. see][]{edelson_2014,smith_2018}
so some caution in comparisons is required. However on the face of it, the PSD slope values of Zw229-15 may be within the distributions of other AGN, but they are on the limits of those distributions. 

\section{Conclusions} 
\label{conclusion}

We present new $\sim$daily optical monitoring of the low mass AGN NGC~4395 over a 3-year period with the LT, LCOGT, Zowada and ZTF. Together with 3s-resolution over a 3 hour period by HiPERCAM \citep{mchardy_2023} and monitoring covering timescales from hours to weeks by TESS \citep{burke_2020}, we are able to measure the optical g'-band PSD over an unprecedented range of almost 7 decades. 
The observed optical PSD is well fitted by a bending PSD model with bend frequency of $3.8^{+4.8}_{-2.6} \times 10^{-6}$\,Hz, high-frequency slope of $2.1^{+0.2}_{-0.4}$ and low-frequency slope of $1^{+0.2}_{-0.2}$.  These parameters, particularly the low-frequency slope, are probably the most accurate yet measured for an AGN optical PSD, but are broadly within the distribution of these parameters as measured from less extensive data by previous researchers.

To test whether the observed PSD can be reproduced by reprocessing of X-rays by an accretion disc we have simulated g'-band lightcurves based on reprocessing of X-ray observations, similar to those observed by \xmm and Swift, from a disc. The g'-band disc response function used is similar to that required to explain the interband lags in NGC~4395 \cite{mchardy_2023}, slightly modified to take account of new \xmm data (Beard et al, in prep).
The PSD of the resulting reprocessed light curve is also well fitted by a bending power law but with a bend frequency almost two decades higher in frequency than that in the observed PSD. The longer bend timescale in the observed PSD is, however,  close to the DRW damping timescale found from analysis of just the TESS data \citep{burke_2020}, although that timescale is towards the bottom of the range of timescales covered by TESS. The PSD slope below the bend, $\alpha_{L}=1.0$, which we are able to constrain well with the 3\,yr timescale lightcurve presented here is, however, steeper than the equivalent PSD slope expected from a damped random walk, i.e. $\alpha_{L}=0$. Our observations are therefore inconsistent with the DRW model for AGN optical variability.  

To test whether a second PSD bend, at the higher frequency implied from disc reprocessing, is compatible with the observed PSD we fitted a model with two bends. The resultant fit does recover both low and high-frequency bends at frequencies similar to those mentioned above, but the fit probability is no improvement on the single bend model. We are able to simulate the observed double-bending power law PSD by the addition of a separate light curve, generated using the formalism of \cite{timmer_1995}, to the synthetic light curve generated from X-ray disc reprocessing. This second lightcurve dominates the low-frequency variability. 

The origin of this second, dominant, source of variability is, at present, unknown. Reprocessing by the BLR, or some mixture of disc wind and BLR \citep{elvis00,hagen24}
which is larger than the disc, will be on longer timescales than disc reprocessing and so may be candidates. The long tails to the reprocessing functions derived by MEMEcho mapping \citep{horne_2004} of NGC~4593 \citep{mchardy_2018} are factors of a few longer than the duration of the disc response function and may represent reprocessing in the BLR. 
In many AGN, eg NGC4593 \citep{cackett_2018}, a clear signature of reprocessing in the BLR is detected as a longer lag in the u-band, approximately a factor of 2 or 3 longer than the expected disc lag.
Frequency-resolved lag analysis of several AGN also shows that the observed lags on timescales of tens of days are longer than can be explained by a standard disc reprocessing model \citep{cackett22,lewin23}.

Thus although the BLR may well contribute power to the PSD on longer timescales than the disc, it is unclear whether the BLR can provide power on timescales 100$\times$ longer than the disc, as required by our observations. This possibility requires further investigation.

An alternative possibility which might more easily explain the long-timescale variations, as it is related to viscous rather than light travel timescales, is inwardly propagating disc accretion rate variations \citep{arevalo_2006_investigating}.  The slow-moving radial disc temperature variations recently found in AGN discs \citep{neustadt22} may be related to this phenomenon, although no claim has yet been made for the origin of these temperature variations.

We conclude that reprocessing of central X-rays by a surrounding accretion disc is consistent with part of the optical variability of NGC~4395 but it cannot explain all of the variability. At least one other source of variations is needed, particularly at low frequencies, but the physical origin of such variations is, as yet, unknown.

\section*{Acknowledgements}

MWJB acknowledges support from the UK Science and Technology Facilities Council (STFC) in the form of studentship ST/S505705/1.
IMcH and DK acknowledge support from STFC from grant ST/V001000/1.
VSD acknowledges support from STFC from grant ST/Z000033/1. EMC acknowledges support from NSF under grant AST-1909199. JVHS acknowledges support from STFC grant ST/V000861/1.
JHK acknowledges grant PID2022-136505NB-I00 funded by
MCIN/AEI/10.13039/501100011033 and also the European Regional Development Fund (MSCA EDUCADO, GA 101119830).

This paper is based on observations by a number of ground-based telescopes, i.e. the Liverpool Telescope (LT), the Las Cumbres Observatory global telescope network (LCOGT), the Gran Telescopio Canarias (GTC), the Zowada observatory, the Zwicky Transient Facility (ZTF), and also on observations by the NASA Transiting Exoplanet Survey Satellite (TESS). 

Both the LT and the GTC are installed at the Spanish Observatorio del Roque de los Muchachos of the Instituto de Astrofisica de Canarias on the island of La Palma. The LT is operated by Liverpool John Moores University with financial support from STFC. LCOGT is a non-profit organisation based in California which runs a network of telescopes around the globe. The Zowada observatory is run by Wayne State University, Michigan. The ZTF is a wide-field survey monitoring facility based at the Palomar Observatory, California.

\section*{Data Availability}

The data underlying this article will be shared on reasonable request to the corresponding author.
%%%%%%%%%%%%%%%%%%%%%%%%%%%%%%%%%%%%%%%%%%%%%%%%%%



\bibliographystyle{mnras}
\bibliography{ref.bib} % if your bibtex file is called example.bib

%%%%%%%%%%%%%%%%%%%%%%%%%%%%%%%%%%%%%%%%%%%%%%%%%%

%%%%%%%%%%%%%%%%% APPENDICES %%%%%%%%%%%%%%%%%%%%%

\appendix

%%%%%%%%%%%%%%%%%%%%%%%%%%%%%%%%%%%%%%%%%%%%%%%%%%

% Don't change these lines
%\bsp	% typesetting comment
\label{lastpage}
\end{document}